\documentclass[10pt,twocolumn,aps,prd,preprintnumbers,showpacs,superscriptaddress,nofootinbib,amsmath,amssymb,floats,floatfix,showkeys,notitlepage,longbibliography,fleqn]{revtex4-2}
\usepackage{comment}
\usepackage{lipsum}
\usepackage{graphicx}
\usepackage{subfigure}
\usepackage{palatino}
\usepackage{sans}
\usepackage{array}
\usepackage[toc,page]{appendix}
\usepackage[normalem]{ulem}
\usepackage{adjustbox}
\usepackage{latexsym}
\usepackage{amsmath}
\usepackage{amssymb}
\usepackage{amsfonts}
\numberwithin{equation}{section}

\usepackage{mathrsfs}
\usepackage{physics}
\usepackage{dcolumn}
\usepackage{bm}
\usepackage{tikz}
\usetikzlibrary{decorations.pathmorphing}
\usepackage{pgfplots}
\pgfplotsset{compat=1.18}
\usepackage{bigints}
\usepackage{array,tabularx,multirow,booktabs}
\usepackage[tracking=true]{microtype}
\usepackage{soul} %for highlighting
\SetTracking{}{500}
\SetTracking{encoding={*}, shape=sc}{40}
\UseRawInputEncoding %for inputenc error%
\allowdisplaybreaks
\usepackage[utf8]{inputenc}
\usepackage{xcolor} % For colored text if needed
\usepackage{babel}
\usepackage{etoolbox}
\usepackage{hyperref}
\hypersetup{colorlinks=true,linkcolor=blue,urlcolor=blue,citecolor=blue}
\usepackage{orcidlink}

\newcommand{\nn}{\nonumber\\}

\begin{document} \sloppy

\title{Phase-plane formulation of weak gravitational deflection in static spherical spacetimes}

\author{Reggie C. Pantig \orcidlink{0000-0002-3101-8591}} 
\email{rcpantig@mapua.edu.ph}
\affiliation{Physics Department, School of Foundational Studies and Education, Map\'ua University, 658 Muralla St., Intramuros, Manila 1002  Philippines.}

\author{Ali \"Ovg\"un \orcidlink{0000-0002-9889-342X}}
\email{ali.ovgun@emu.edu.tr}
\affiliation{Physics Department, Eastern Mediterranean University, Famagusta, 99628 North Cyprus via Mersin 10, T\"urkiye}

\begin{abstract}
This paper develops a phase-plane formulation of gravitational light deflection by static and spherically symmetric black holes, with Schwarzschild spacetime as the principal case. Instead of perturbing the null trajectory and locating the displaced outgoing asymptote, we represent the orbit through an amplitude and an intrinsic phase. The bending angle then follows from the excess physical azimuth accumulated while the intrinsic phase advances between two fixed asymptotic endpoints. For Schwarzschild spacetime, the exact radial first integral reduces the amplitude evolution to a cubic algebraic relation. Its physical branch generates the local phase factor through a single inverse algebraic map. Lagrange inversion then yields an explicit all-order weak-deflection coefficient formula in powers of the invariant ratio \(M/b\). Each coefficient separates into an algebraic phase contribution and a universal trigonometric moment, which explains the alternating rational and \(\pi\)-dependent structure of the Schwarzschild series. Independent comparison with the exact radial scattering integral and conventional orbit perturbation reproduces the standard weak-bending coefficients. The same phase framework extends to general static spherical geometries, while the Schwarzschild cubic represents an especially simple member of a broader algebraic class. The branch singularity of the phase map also coincides with the critical photon orbit and governs the convergence of the weak-deflection expansion.
\end{abstract}


\maketitle

\section{Introduction}
\label{sec1}
Gravitational deflection of light occupies a special place in general relativity because it connects the geometry of spacetime directly with an observable propagation effect. Einstein obtained the relativistic light-bending prediction within the general theory of relativity \cite{Einstein:1916vd}, and the eclipse observations reported by Dyson, Eddington, and Davidson soon placed this prediction among the earliest empirical successes of the theory \cite{Dyson:1920cwa}. Einstein later emphasized the lens-like action of gravitating bodies \cite{Einstein:1936llh}. The subject subsequently developed from solar-system light propagation into the broader theory of gravitational lensing, where compact objects provide an especially clean setting because the exterior geometry can often be described exactly. For the Schwarzschild spacetime, Darwin's analyses already exposed the nontrivial behavior of null trajectories near the unstable circular photon orbit \cite{Darwin_1959,Darwin_1961}. These early developments established two regimes that remain fundamental today. Weak deflection describes rays passing far outside the gravitational radius, while strong deflection describes rays approaching the critical photon orbit and accumulating large angular changes before escaping.

We focus here on the nonrotating case because static spherical symmetry separates the essential mathematics of gravitational scattering from complications introduced by frame dragging and nonspherical orbital motion. Even within this restricted setting, the lensing problem contains several distinct mathematical layers. One may study the null geodesic itself, the mapping between source and image directions, the optical expansion and shear, the radial scattering integral, or the geometry of the optical manifold. Dyer used optical scalars to analyze spherical lenses beyond the elementary Einstein formula \cite{Dyer_1977}. Frittelli and Newman developed exact lensing relations without relying on the usual background-space construction \cite{Frittelli:1998hr}, while Frittelli, Kling, and Newman formulated Schwarzschild lensing directly from the spacetime geometry \cite{Frittelli:1999yf}. Virbhadra and Ellis examined the full Schwarzschild lensing map and its relativistic images \cite{Virbhadra:1999nm}. Perlick later formulated an exact lens equation for static spherical geometries and reviewed gravitational lensing from a fully spacetime-based viewpoint \cite{Perlick:2003vg,Perlick:2004tq}. These works make clear that the familiar weak lens equation represents only one approximation within a much richer relativistic scattering problem.

The weak-deflection angle itself has received sustained attention because higher-order terms carry physical information that the leading Einstein contribution cannot contain. Epstein and Shapiro calculated the post-post-Newtonian light deflection by the Sun \cite{Epstein:1980dw}, while Fischbach and Freeman independently derived the second-order contribution \cite{Fischbach:1980su}. Richter and Matzner developed a parametrized post-Newtonian treatment at the same level and subsequently extended the analysis to general three-dimensional photon trajectories \cite{Richter:1982zz,Richter:1982zza}. Bodenner and Will later used the second-order Schwarzschild problem to emphasize an important conceptual point. Intermediate expressions written in terms of a coordinate radius of closest approach can differ among radial coordinate systems, while expressions written in terms of measurable asymptotic quantities agree \cite{Bodenner_2003}. This distinction motivates the use of the invariant impact parameter when we organize weak scattering.

A systematic invariant framework became especially useful once gravitational lensing began to serve as a possible probe of deviations from Schwarzschild geometry. Sereno discussed weak lensing within general metric theories of gravity \cite{Sereno:2003tk}. Keeton and Petters constructed a general perturbative treatment for static spherical compact objects and organized lensing observables through invariant weak-field coefficients \cite{Keeton:2005jd}. They subsequently developed the post-post-Newtonian consequences of the same framework \cite{Keeton:2006sa}. Iyer and Petters studied the Schwarzschild bending angle across weak and strong regimes using an invariant impact-parameter description \cite{Iyer:2006cn}. Their formulation is particularly relevant to the present work because it demonstrates why a physically meaningful expansion parameter should remain tied to asymptotic scattering quantities rather than to a coordinate-dependent turning radius.

Exact lens equations and improved approximate lens equations address a related but logically different issue. They determine how a bending law enters the observable source-image geometry once the ray has propagated between lens, source, and observer. Bozza compared commonly used approximate lens equations and proposed a more accurate form beyond the simplest small-angle approximation \cite{Bozza:2008ev}. Zschocke developed a generalized weak-field lens equation that retains finite source and observer distances in Schwarzschild geometry \cite{Zschocke:2011mm}. These developments complement the present problem rather than replace it. We concentrate on the bending angle itself and ask whether its perturbative calculation can be reorganized before one inserts it into any particular lens equation.

The standard geodesic route remains the most direct approach to weak Schwarzschild bending. One obtains a nonlinear orbit equation for the inverse radius, expands the trajectory in a small gravitational parameter, and extracts the deflection from the shifted outgoing asymptote. This method is conceptually straightforward, but higher orders require increasingly complicated orbit corrections and repeated expansions of the asymptotic root. Several authors have therefore sought alternative perturbative organizations. Amore and Arceo introduced rapidly convergent analytical approximations for light deflection in general static spherical metrics \cite{Amore:2006pi}, and Amore, Arceo, and Fern\'andez extended that strategy to higher orders \cite{Amore:2006xp}. Mak, Leung, and Harko applied the Laplace--Adomian decomposition method to Schwarzschild null motion and obtained successive approximations to the bending angle \cite{Mak:2018hcn}. Rodriguez and Mar\'in used the Lindstedt--Poincar\'e method and Pad\'e approximants to generate high-order Schwarzschild corrections \cite{Rodriguez:2017pky}. Jia developed a systematic perturbative treatment for weak deflection in general static spherical geometries and generated Schwarzschild results to high order \cite{Jia:2020xbc}. Sasaki and Suzuki approached the same Schwarzschild bending problem through an inhomogeneous Picard--Fuchs equation and obtained analytic weak and strong expansions from the resulting integral structure \cite{Sasaki:2020kmg}. The diversity of these approaches shows that the weak-deflection series is not tied to one perturbative technique.

Optical geometry provides a fundamentally different viewpoint. Gibbons and Werner showed that the Gauss--Bonnet theorem applied to the optical metric converts weak gravitational bending into a statement involving Gaussian curvature \cite{Gibbons:2008rj}. Gibbons and Warnick further examined universal features of black hole optical geometries \cite{Gibbons:2008hb}. Ishihara, Suzuki, Ono, Kitamura, and Asada used the same geometric framework to define light deflection when the source and observer remain at finite distances \cite{Ishihara:2016sfv}. Ishihara, Suzuki, Ono, and Asada then extended this construction toward the strong-deflection regime \cite{Ishihara:2016vdc}. Crisnejo and Gallo adapted the Gauss--Bonnet approach to weak lensing in dispersive environments and to massive-particle trajectories \cite{Crisnejo:2018uyn}, while Crisnejo, Gallo, and Villanueva developed related curvature-based expressions for more general propagation settings \cite{Crisnejo:2019xtp}. Ono and Asada reviewed the finite-distance formulation and clarified its geometrical interpretation \cite{Ono:2019hkw}. Li and Zhou established the equivalence between the Gibbons--Werner construction and the standard geodesic method under asymptotically flat conditions \cite{Li:2019mqw}. More recently, Pantig and \"Ovg\"un developed a reference-based curvature formulation for finite-distance weak lensing in static spherical spacetimes \cite{Pantig:2026xjj}. These studies show that a bending angle obtained from the geodesic equation can often be reorganized into a geometrically different quantity without changing the observable scattering angle.

The strong-deflection literature supplies another important perspective because it identifies the analytic boundary that any weak expansion eventually encounters. Claudel, Virbhadra, and Ellis formulated the general notion of photon surfaces \cite{Claudel:2000yi}. Bozza, Capozziello, Iovane, and Scarpetta developed an analytic strong-field treatment of Schwarzschild lensing \cite{Bozza:2001xd}, and Bozza subsequently derived the logarithmic strong-deflection structure for general static spherical spacetimes \cite{Bozza:2002zj}. Eiroa, Romero, and Torres studied the corresponding behavior for the Reissner--Nordstr\"om geometry \cite{Eiroa:2002mk}. Bozza and Scarpetta incorporated arbitrary source distances into strong-deflection lensing \cite{Bozza:2007gt}. Bozza later reviewed black hole lensing across its principal regimes \cite{Bozza:2010xqn}. Tsukamoto refined the strong-deflection expansion for general asymptotically flat static spherical geometries \cite{Tsukamoto:2016jzh}, while Jia and Huang developed another perturbative description near critical scattering \cite{Jia:2020qzt}. Recent work has continued the study of null rays near marginally unstable photon spheres in general static spherical spacetimes \cite{Tsukamoto:2025hbz}. Although our main concern lies in weak deflection, these studies matter because the nearest critical orbit determines where a weak expansion can cease to remain analytic.

Finite-distance and related propagation effects further demonstrate the usefulness of methods that separate local orbital evolution from the final scattering observable. Zschocke's generalized Schwarzschild lens equation already exhibits how finite radial positions modify the usual asymptotic picture \cite{Zschocke:2011mm}. Ishihara and collaborators supplied a geometric definition suitable for finite source and receiver positions \cite{Ishihara:2016sfv,Ishihara:2016vdc}. Ono and Asada summarized the resulting framework \cite{Ono:2019hkw}. Liu and Jia developed systematic weak-field expansions for propagation times in arbitrary static spherical geometries \cite{Liu:2020wcu}. These developments suggest that a useful formulation of bending should make clear which structures arise from the local orbit and which arise from the choice of asymptotic or finite endpoints.

Despite this extensive literature, most analytical calculations of the static spherical bending angle begin from one of a small number of mathematical objects. We either perturb the null trajectory, expand a radial scattering integral, manipulate an exact lens equation, or integrate a curvature quantity associated with the optical geometry \cite{Epstein:1980dw,Keeton:2005jd,Iyer:2006cn,Gibbons:2008rj,Jia:2020xbc,Sasaki:2020kmg}. Each route ultimately measures the mismatch between the incoming and outgoing propagation directions. The observable therefore has the character of an accumulated angular change, but conventional orbit perturbation usually recovers that change indirectly by determining where the perturbed trajectory returns to infinity.

Phase-amplitude methods offer a natural language for separating an oscillatory trajectory into radial motion in phase space and angular advance. Such descriptions have a long history in nonlinear dynamics and continue to provide useful reductions of nonlinear oscillator equations. Mayol, Toral, and Mirasso derived amplitude equations for nonlinear oscillators under general forcing \cite{Mayol_2004}, while Wedgwood, Lin, Thul, and Coombes developed phase-amplitude descriptions that retain information lost in a phase-only treatment \cite{Wedgwood_2013}. We do not transfer those physical models to gravitational lensing. We instead borrow the elementary geometric principle that a second-order nonlinear oscillator can sometimes be understood more transparently by following its phase-plane radius and phase rather than its Cartesian orbit variable.

This observation motivates the present work. The Schwarzschild null-orbit equation has the form of a weakly nonlinear oscillator when we use the azimuthal angle as the independent variable and the inverse radius as the orbital variable. We therefore ask whether the gravitational bending angle can be defined directly through the accumulated difference between the physical azimuth and an intrinsic phase-plane angle. Such a formulation would replace the moving outgoing root of the conventional perturbative orbit by fixed phase endpoints. It would also separate the question of how the orbit changes its phase-plane amplitude from the question of how much physical azimuth accumulates during one scattering passage.

We pursue this possibility first for Schwarzschild spacetime because its exact radial first integral, invariant impact parameter, and critical photon orbit are known without approximation. Our aim is methodological rather than phenomenological. We seek an exact amplitude-phase description of the null orbit, an observable-first expression for the asymptotic deflection angle, and a systematic weak-field expansion built from that description. We then ask which parts of the construction depend on the special Schwarzschild nonlinearity and which survive for a general static spherical metric. Throughout the analysis, we use the standard geodesic and radial-integral formulations as independent reference descriptions rather than modifying the physical definition of the bending angle.

The paper is organized as follows. Section \ref{sec2} formulates the invariant Schwarzschild null-scattering problem in terms of the asymptotic impact parameter. Section \ref{sec3} introduces the exact phase-plane variables and formulates the bending angle through a fixed intrinsic phase interval. Section \ref{sec4} examines the algebraic structure of the Schwarzschild amplitude equation. Section \ref{sec5} develops the corresponding all-order weak-deflection expansion. Section \ref{sec6} compares the formulation with the exact radial integral and conventional orbit perturbation. Section \ref{sec7} extends the phase construction to general static spherical geometries. Section \ref{sec8} discusses the domain of validity, critical scattering, convergence, and mathematical limitations. Section \ref{sec9} summarizes the conclusions and outlines further directions.

\section{Invariant Schwarzschild Null Scattering Problem}
\label{sec2}
We formulate the Schwarzschild light-scattering problem in terms of quantities that retain a direct asymptotic physical meaning. We work in geometrized units with \(G=c=1\), take \(M>0\), and describe the exterior spacetime by
\begin{align}
ds^{2}
&=
-f(r)\,dt^{2}
+
\frac{dr^{2}}{f(r)}
+
r^{2}
\left(
d\theta^{2}
+
\sin^{2}\theta\,d\phi^{2}
\right),
\nn
f(r)&=1-\frac{2M}{r}.
\end{align}

Spherical symmetry allows us to rotate any null geodesic into an equatorial plane without changing its physical content. We therefore choose \(\theta=\frac{\pi}{2}\), and \(\dot{\theta}=0,\) where a dot denotes differentiation with respect to an affine parameter \(\lambda\). The null condition then reads
\begin{equation}
-f(r)\dot t^{\,2}
+
\frac{\dot r^{\,2}}{f(r)}
+
r^{2}\dot\phi^{\,2}
=
0.
\label{2.3}
\end{equation}

Stationarity and axial symmetry generate two conserved quantities. We normalize the timelike Killing vector by its standard asymptotic Minkowski normalization and write
\begin{align}
E
=
f(r)\dot t,
\qquad
L
=
r^{2}\dot\phi.
\label{2.4}
\end{align}
Here \(E\) and \(L\) represent the conserved energy and angular momentum associated with the chosen affine parametrization. A constant rescaling of \(\lambda\) rescales both quantities by the same factor. Their ratio therefore remains unchanged.

We define the asymptotic impact parameter by
\begin{equation}
b=\frac{L}{E}.
\label{2.5}
\end{equation}
For an asymptotically flat null scattering orbit, \(b\) coincides with the usual perpendicular separation between the corresponding undeflected ray and the center of symmetry at infinity. We will organize the weak-field expansion through \(M/b\), rather than through the coordinate distance of closest approach. This choice prevents an unnecessary dependence on an intermediate radial turning point.

Substitution of Eq. \eqref{2.4} into the null condition in Eq. \eqref{2.3} gives
\begin{equation}
\dot r^{\,2}
=
E^{2}
-
f(r)\frac{L^{2}}{r^{2}}.
\label{2.6}
\end{equation}
We may regard Eq. \eqref{2.6} as the radial energy equation for the photon. The second term contains the centrifugal contribution together with the Schwarzschild modification. A scattering orbit starts at \(r\rightarrow\infty\), reaches one finite radial minimum, and returns to \(r\rightarrow\infty\).

We introduce the inverse curvature radius \(u=1/r\). Since Eq. \eqref{2.4} gives \(\dot\phi=L/r^{2}\), differentiation along the orbit yields
\begin{equation}
\frac{du}{d\phi}
=
-\frac{\dot r}{L}.
\label{2.8}
\end{equation}
Combining Eqs. \eqref{2.6} and \eqref{2.8}, and using Eq. \eqref{2.5}, we obtain the exact first integral
\begin{equation}
\left(
\frac{du}{d\phi}
\right)^{2}
=
\frac{1}{b^{2}}
-u^{2}
+2Mu^{3}.
\label{2.9}
\end{equation}

The three terms in Eq. \eqref{2.9} have the same dimension \(L^{-2}\). The first term fixes the asymptotic angular momentum scale, the second term gives the flat-space centrifugal behavior, and the cubic term contains the Schwarzschild correction. The cubic term is ultimately responsible for both weak gravitational bending and the unstable photon orbit.

Differentiating Eq. \eqref{2.9} with respect to \(\phi\) gives
\begin{equation}
2u'u''
=
\left(
-2u+6Mu^{2}
\right)u'.
\end{equation}
Away from a radial turning point we divide by \(2u'\) and obtain
\begin{equation}
u''+u
=
3Mu^{2}.
\label{2.11}
\end{equation}
Both sides of Eq. \eqref{2.11} remain regular at the turning point, so continuity extends the relation through the point where \(u'=0\). We therefore use Eq. \eqref{2.11} along the complete scattering orbit. An independent symbolic differentiation of Eq. \eqref{2.9} confirms the sign and coefficient of the nonlinear term.

We now remove the remaining dimensions by defining
\(z=bu,\) and \(\epsilon= M / b\). The quantity \(\epsilon\) provides the natural invariant weak-scattering parameter. Equation \eqref{2.9} becomes
\begin{equation}
z'^{\,2}
=
1-z^{2}+2\epsilon z^{3},
\label{2.13}
\end{equation}
while Eq. \eqref{2.11} becomes
\begin{equation}
z''+z
=
3\epsilon z^{2}.
\label{2.14}
\end{equation}
We will regard Eq. \eqref{2.14}, together with its first integral in Eq. \eqref{2.13}, as the basic dynamical system. The inverse-radius variable \(z\) depends on our choice of Schwarzschild curvature radius, but the weak parameter \(\epsilon\) and the asymptotic scattering angle do not depend on the arbitrary affine normalization.

The first integral also identifies the scattering domain. Let \(z_{0}\) denote the value of \(z\) at the radial turning point. Since \(z'=0\) there, Eq. \eqref{2.13} gives
\begin{equation}
1-z_{0}^{2}
+
2\epsilon z_{0}^{3}
=
0.
\label{2.15}
\end{equation}
A returning null orbit requires an accessible positive root of this equation. The limiting orbit occurs when the relevant root becomes degenerate. We impose the two conditions
\begin{align}
1-z_{\mathrm{c}}^{2}
+
2\epsilon_{\mathrm{c}}z_{\mathrm{c}}^{3}
=
0,
\nn
-2z_{\mathrm{c}}
+
6\epsilon_{\mathrm{c}}z_{\mathrm{c}}^{2}
=
0.
\end{align}
The nonzero solution gives
\begin{align}
z_{\mathrm{c}}
=
\sqrt{3},
\qquad
\epsilon_{\mathrm{c}}
=
\frac{1}{3\sqrt{3}},
\qquad
b_{\mathrm{c}}
=
3\sqrt{3}\,M.
\end{align}
The corresponding curvature radius is \(r=3M\), which is the Schwarzschild photon-sphere radius. We restrict the present analysis to returning trajectories with
\begin{align}
b>b_{\mathrm{c}},
\qquad
0<\epsilon<\frac{1}{3\sqrt{3}}.
\end{align}
Weak deflection occupies the stronger regime \(\epsilon\ll1\).
The distinction matters because a scattering orbit may exist even when a low-order expansion in \(M/b\) no longer provides an accurate representation.

We next fix the angular convention required to define the deflection angle without ambiguity. We place the incoming asymptote at
\begin{equation}
\phi_{\mathrm{in}}=0.
\label{2.20}
\end{equation}
As \(r\rightarrow\infty\), we have \(z\rightarrow0\). Equation \eqref{2.13} then gives \(z'^{\,2}\rightarrow1.\)
Along the incoming branch, \(r\) decreases while \(u\) and \(z\) increase. We therefore select the positive sign and impose
\begin{align}
z(0)=0,
\qquad
z'(0)=1.
\label{2.22}
\end{align}
These conditions remove any remaining freedom associated with translating or reversing the azimuthal parameter.

The physical meaning becomes transparent in the zero-mass limit. Setting \(\epsilon=0\) in Eq. \eqref{2.14} gives \(z''+z=0.\)
The conditions in Eq. \eqref{2.22} select \(z_{\mathrm{flat}}(\phi)=\sin\phi.\)
The ray begins at infinity when \(\phi=0\), reaches its minimum radius at \(\phi=\pi/2\), and returns to infinity at \(\phi=\pi\). The total change in azimuth equals \(\pi\), so the flat trajectory has zero deflection.

For \(M>0\), the nonlinear term in Eq. \eqref{2.14} shifts the second asymptotic zero beyond \(\pi\). We denote the outgoing angular position by
\begin{equation}
\phi_{\mathrm{out}}
=
\pi+\alpha.
\label{2.25}
\end{equation}
The asymptotic condition therefore takes the form
\begin{align}
z(\pi+\alpha)=0,
\qquad
z'(\pi+\alpha)<0.
\label{2.26}
\end{align}
We define the positive Schwarzschild bending angle as
\begin{equation}
\alpha
=
\phi_{\mathrm{out}}-\phi_{\mathrm{in}}-\pi.
\label{2.27}
\end{equation}
With the convention in Eq. \eqref{2.20}, Eq. \eqref{2.27} reduces directly to the parameter \(\alpha\) appearing in Eq. \eqref{2.25}.

The definition in Eq. \eqref{2.27} has a simple asymptotic meaning. Far from the black hole, the Schwarzschild angular coordinate approaches the ordinary polar angle of Minkowski space. The difference between the actual angular sweep and the flat value \(\pi\) therefore measures the physical change between the incoming and outgoing propagation directions. We do not need to assign physical significance to the coordinate location of the turning point.

Conventional perturbation theory would now expand the solution of Eq. \eqref{2.14} as
\begin{equation}
z(\phi)
=
z_{0}(\phi)
+
\epsilon z_{1}(\phi)
+
\epsilon^{2}z_{2}(\phi)
+
\mathcal{O}(\epsilon^{3})
\end{equation}
and then solve Eq. \eqref{2.26} perturbatively for the displaced outgoing zero. At leading order, the result takes the familiar form
\begin{equation}
\alpha
=
\frac{4M}{b}
+
\mathcal{O}\!\left(\frac{M^{2}}{b^{2}}\right).
\end{equation}
Although this procedure works, it determines the observable indirectly. The orbit changes order by order, and the location at which we impose the outgoing asymptotic condition changes with it.

We instead retain the exact orbit equation \eqref{2.14} with the initial conditions \eqref{2.22} and reinterpret its scattering geometry before carrying out any expansion. The essential mathematical question is then not how far the second zero of \(z(\phi)\) moves from \(\pi\). We ask how gravity changes the physical azimuth required for the orbit to execute one fixed half-cycle in its phase plane. That reformulation preserves the same invariant observable in Eq. \eqref{2.27} while changing the object on which the weak-field expansion acts.

\section{Exact Phase-Plane Formulation}
\label{sec3}
We now reformulate the Schwarzschild scattering problem so that the deflection angle appears directly as an accumulated phase shift. Our starting point is the exact dimensionless orbit equation \eqref{2.14} with the conditions \eqref{2.22}. The conventional perturbative treatment regards \(z(\phi)\) as the primary quantity and determines the deflection angle from the displacement of its outgoing zero. We instead regard the pair \((z,z')\) as a trajectory in a two-dimensional phase plane and follow its angular advance.

We introduce an amplitude \(A(\phi)\) and a phase \(\Theta(\phi)\) through
\begin{align}
z
=
A\sin\Theta,
\qquad
z'
=
A\cos\Theta.
\label{3.2}
\end{align}
The amplitude follows directly from the phase-plane radius
\begin{equation}
A
=
\sqrt{
z^{2}
+
z'^{\,2}
}.
\end{equation}
For the scattering solutions considered here, \(A\) remains positive. The first integral in Sec. \ref{sec2} prevents \(z\) and \(z'\) from vanishing simultaneously. We may therefore define \(\Theta\) continuously along the complete orbit.

At the incoming asymptote, the initial conditions in Eq. \eqref{2.22} give \(A(0)=1.\)
The same conditions select \(\Theta(0)=0.\)
Our choice of phase convention makes the flat-space solution especially simple. When \(\epsilon=0\), we have \(A=1\) and \(\Theta=\phi\), so Eq. \eqref{3.2} reduces immediately to \(z=\sin\phi\).

\begin{figure}[t]
\centering
\includegraphics[width=\columnwidth]{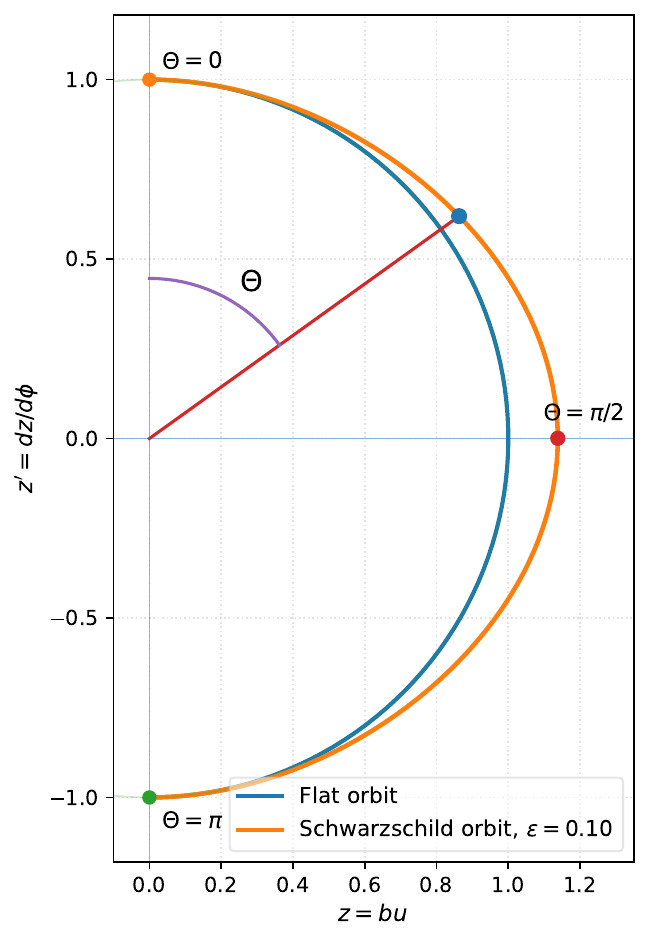}
\caption{Phase portrait of flat and Schwarzschild null scattering in the \((z,z')\) plane, with the amplitude \(A\) and intrinsic phase \(\Theta\) indicated geometrically.}
\label{fig1}
\end{figure}
The phase portrait makes the change of variables geometrically transparent. The flat null orbit traces the unit semicircle in the \((z,z')\) plane, whereas the Schwarzschild trajectory moves outward from that reference curve as the photon approaches the black hole and returns to the same asymptotic phase-plane radius after closest approach. At any point on the trajectory, the phase-space radius gives \(A=\sqrt{z^{2}+z'^{\,2}}\), while the angle measured from the positive \(z'\) direction gives \(\Theta\). The incoming asymptote, closest approach, and outgoing asymptote correspond respectively to \(\Theta=0\), \(\Theta=\pi/2\), and \(\Theta=\pi\). Fig. \ref{fig1} therefore shows geometrically why gravitational bending can be studied through the evolution of an intrinsic phase rather than only through the radial orbit \(z(\phi)\).

We derive the exact evolution equations for \(A\) and \(\Theta\) before introducing any weak-field approximation. Differentiating the first relation in Eq. \eqref{3.2} with respect to \(\phi\) gives
\begin{equation}
z'
=
A'\sin\Theta
+
A\Theta'\cos\Theta.
\label{3.6}
\end{equation}
Equating this result with the second relation in Eq. \eqref{3.2}, we obtain
\begin{equation}
A'\sin\Theta
+
A\left(
\Theta'-1
\right)
\cos\Theta
=
0.
\label{3.8}
\end{equation}

A second differentiation starts from \(z'=A\cos\Theta\) and gives
\begin{equation}
z''
=
A'\cos\Theta
-
A\Theta'\sin\Theta.
\label{3.9}
\end{equation}
We substitute Eqs. \eqref{3.2} and \eqref{3.9} into the exact Schwarzschild orbit equation \eqref{2.14}. The result is
\begin{equation}
A'\cos\Theta
+
A\left(
1-\Theta'
\right)
\sin\Theta
=
3\epsilon
A^{2}
\sin^{2}\Theta.
\label{3.10}
\end{equation}

We now solve Eqs. \eqref{3.8} and \eqref{3.10} for the two unknown derivatives \(A'\) and \(\Theta'\). Multiplying Eq. \eqref{3.8} by \(\sin\Theta\), multiplying Eq. \eqref{3.10} by \(\cos\Theta\), and adding the resulting expressions gives
\begin{equation}
A'
\left(
\sin^{2}\Theta
+
\cos^{2}\Theta
\right)
=
3\epsilon
A^{2}
\sin^{2}\Theta
\cos\Theta.
\end{equation}
We therefore obtain the exact amplitude equation
\begin{equation}
A'
=
3\epsilon
A^{2}
\sin^{2}\Theta
\cos\Theta.
\label{3.12}
\end{equation}

We determine the phase equation by multiplying Eq. \eqref{3.8} by \(\cos\Theta\), multiplying Eq. \eqref{3.10} by \(\sin\Theta\), and subtracting the first expression from the second. This operation gives
\begin{equation}
A\left(
1-\Theta'
\right)
=
3\epsilon
A^{2}
\sin^{3}\Theta.
\end{equation}
Since \(A>0\), we find
\begin{equation}
\Theta'
=
1
-
3\epsilon
A\sin^{3}\Theta.
\label{3.14}
\end{equation}
Equations \eqref{3.12} and \eqref{3.14} exactly reproduce the Schwarzschild null dynamics in amplitude-phase variables. We have not expanded either \(A\) or \(\Theta\) in powers of \(\epsilon\).

The physical interpretation of Eq. \eqref{3.14} is immediate in the zero-mass limit. Setting \(\epsilon=0\) gives \(\Theta'=1,\)
which reproduces \(\Theta=\phi\) under the initial condition \(\Theta(0)=0.\) Schwarzschild gravity modifies this unit phase-advance rate through the term \(3\epsilon A\sin^{3}\Theta\). We can therefore regard the bending angle as the accumulated difference between the physical azimuthal advance and the corresponding phase advance.

Before using this interpretation, we establish the behavior of \(\Theta\) along the full returning orbit. At the radial turning point, \(z'=0\). Equation \eqref{3.2} then requires \(\cos\Theta_{0}=0.\)
The branch connected continuously to the incoming value \(\Theta=0\) reaches \(\Theta_{0}=\frac{\pi}{2}.\)
At this point, \(z_{0}=A_{0},\)
where \(z_{0}\) denotes the dimensionless inverse radius at closest approach.

For a returning Schwarzschild ray with \(b>b_{\mathrm{c}}\), the turning point lies on the outer branch
\begin{equation}
1<z_{0}<\sqrt{3}.
\label{3.19}
\end{equation}
Using the turning-point equation \eqref{2.15}, the combination that appears naturally in the phase equation is
\begin{equation}
3\epsilon z_{0}
=
\frac{3}{2}
\left(
1-\frac{1}{z_{0}^{2}}
\right).
\end{equation}
The bound in Eq. \eqref{3.19} therefore implies
\begin{equation}
0
<
3\epsilon z_{0}
<
1.
\label{3.22}
\end{equation}

We can use this inequality to prove that the phase remains monotonic along every returning orbit outside the critical trajectory. From Eq. \eqref{3.2}, we have
\begin{equation}
A\sin^{3}\Theta
=
z\sin^{2}\Theta.
\end{equation}
Throughout the scattering orbit,
\begin{equation}
0
\leq
z\sin^{2}\Theta
\leq
z
\leq
z_{0}.
\label{3.24}
\end{equation}
Combining Eqs. \eqref{3.14}, \eqref{3.22}, and \eqref{3.24}, we obtain
\begin{equation}
\Theta'
\geq
1-3\epsilon z_{0}
>
0.
\label{3.25}
\end{equation}
Hence \(\Theta\) increases strictly from the incoming asymptote to the outgoing asymptote. The phase variable therefore provides a valid global parameter for the entire returning null trajectory.

The outgoing asymptote satisfies \(z=0\) and \(z'<0\). Since \(A>0\), Eq. \eqref{3.2} requires \(\sin\Theta_{\mathrm{out}}=0,\) and \(\cos\Theta_{\mathrm{out}}<0.\)
Monotonicity and continuity then select \(\Theta_{\mathrm{out}} = \pi\)
The complete scattering trajectory consequently spans the fixed phase interval \(0\leq\Theta\leq\pi.\) Gravity changes the physical value of \(\phi\) required to cross this interval, but it does not move either phase endpoint.

The amplitude equation also acquires a transparent geometrical meaning. Equation \eqref{3.12} gives \(A'>0\), \(0<\Theta<\frac{\pi}{2},\)
while \(A'<0\), and \(\frac{\pi}{2}<\Theta<\pi.\)
Thus the phase-plane radius increases from its asymptotic value \(A=1\), reaches its maximum at closest approach, and decreases again on the outgoing branch. At either asymptote, \(z=0\), and the first integral gives \(z'^{\,2}=1\). Hence, \(A_{\mathrm{in}}=A_{\mathrm{out}}=1\)
The gravitational interaction deforms the phase-plane trajectory away from the unit circle while preserving the same asymptotic phase-plane radius.

Since \(\Theta\) varies monotonically, we may invert Eq. \eqref{3.14} and use \(\Theta\) itself as the independent variable. We obtain
\begin{equation}
\frac{d\phi}{d\Theta}
=
\frac{1}
{
1
-
3\epsilon
A(\Theta)
\sin^{3}\Theta
}.
\label{3.32}
\end{equation}
The total physical azimuth accumulated between the two asymptotes is therefore
\begin{equation}
\Delta\phi
=
\int_{0}^{\pi}
\frac{
d\Theta
}
{
1
-
3\epsilon
A(\Theta)
\sin^{3}\Theta
}.
\label{3.33}
\end{equation}

In flat spacetime, the same phase advance from \(0\) to \(\pi\) requires an azimuthal change equal to \(\pi\). Thus \(\alpha=\Delta\phi-\pi\), and substitution of Eq. \eqref{3.33} gives our basic phase-plane representation
\begin{equation}
\alpha
=
\int_{0}^{\pi}
\left[
\frac{1}
{
1
-
3\epsilon
A(\Theta)
\sin^{3}\Theta
}
-
1
\right]
d\Theta.
\label{3.35}
\end{equation}

We can display the gravitational contribution more explicitly by combining the two terms in the integrand. We find
\begin{equation}
\alpha
=
3\epsilon
\int_{0}^{\pi}
\frac{
A(\Theta)\sin^{3}\Theta
}
{
1
-
3\epsilon
A(\Theta)\sin^{3}\Theta
}
\,d\Theta.
\end{equation}
For every returning Schwarzschild trajectory considered here, Eq. \eqref{3.25} keeps the denominator positive. The remaining factors in the integrand are nonnegative for \(0\leq\Theta\leq\pi\). We therefore recover directly \(\alpha>0\) for \(M>0\), which expresses the attractive Schwarzschild deflection within the phase formulation itself.

The central change in viewpoint appears in Eq. \eqref{3.35}. The conventional orbit method fixes the incoming value of \(\phi\) and searches for a gravitationally displaced outgoing zero at \(\phi=\pi+\alpha\). Our formulation instead fixes the phase endpoints exactly at \(0\) and \(\pi\). The unknown bending angle resides entirely in the mismatch between the rates \(d\phi\) and \(d\Theta\).

We may express the same statement locally by defining the gravitational phase lag per unit intrinsic phase as
\begin{equation}
\mathcal{P}(\Theta)
=
\frac{d\phi}{d\Theta}
-
1.
\end{equation}
Equation \eqref{3.32} gives
\begin{equation}
\mathcal{P}(\Theta)
=
\frac{
3\epsilon
A(\Theta)\sin^{3}\Theta
}
{
1
-
3\epsilon
A(\Theta)\sin^{3}\Theta
}.
\end{equation}
The total bending angle is simply
\begin{equation}
\alpha
=
\int_{0}^{\pi}
\mathcal{P}(\Theta)\,d\Theta.
\end{equation}
We thus identify the observable deflection angle with an accumulated phase lag over one fixed half-cycle of the null orbit.

\begin{figure*}[t]
\centering
\includegraphics[width=0.48\textwidth]{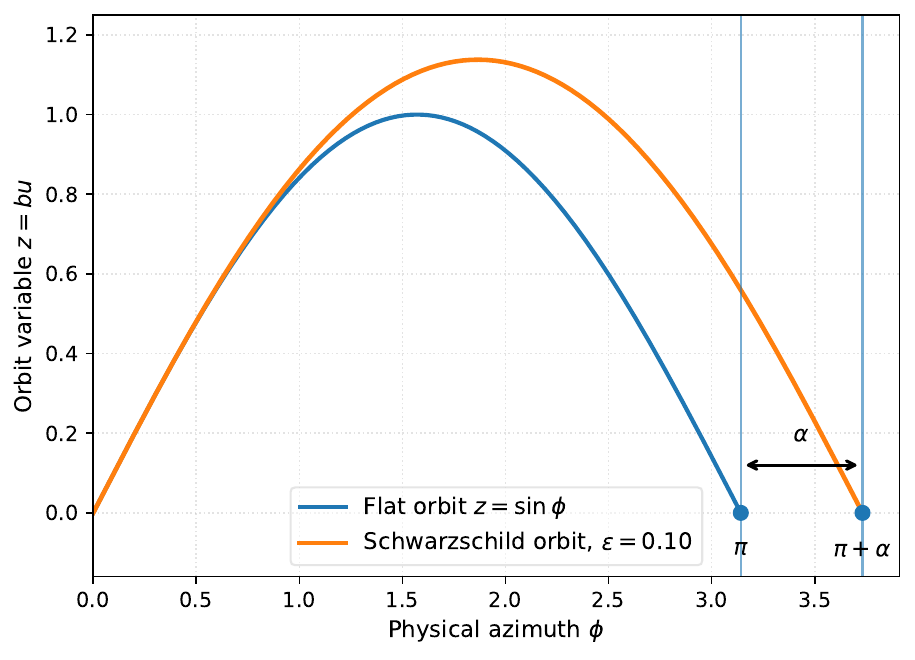}
\includegraphics[width=0.48\textwidth]{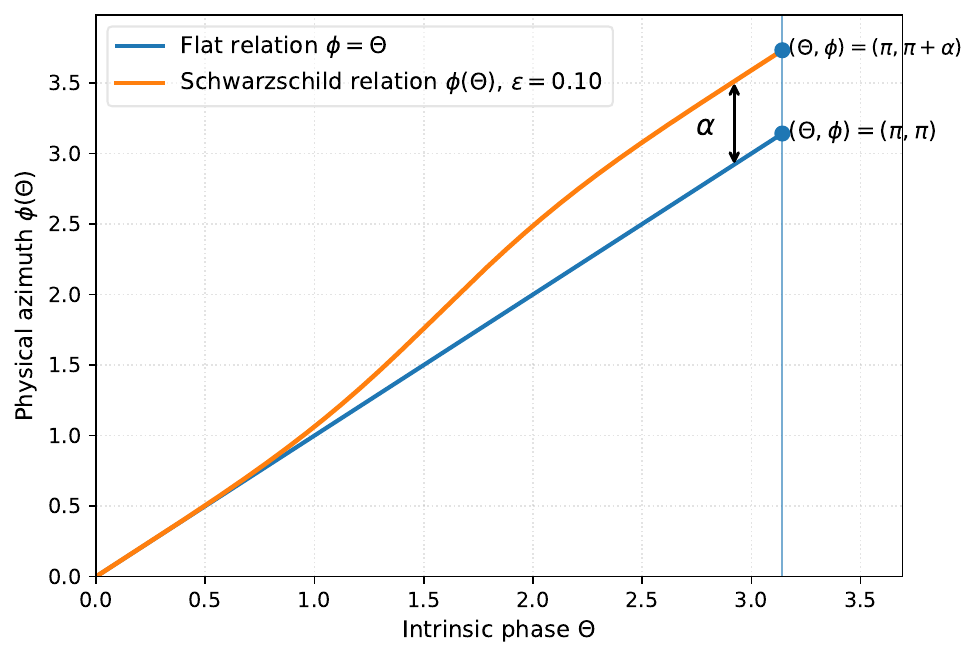}
\caption{Equivalent representations of Schwarzschild bending. The left panel shows the displaced outgoing root, while the right panel keeps the intrinsic phase interval fixed and represents \(\alpha\) as excess physical azimuth.}
\label{fig2}
\end{figure*}
A direct comparison of the two descriptions reveals that they encode precisely the same asymptotic scattering angle while assigning the gravitational correction to different quantities. The left panel follows the conventional orbit picture, where the outgoing zero moves from its flat-space position \(\phi=\pi\) to \(\phi=\pi+\alpha\). See Fig. \ref{fig2}. The right panel instead fixes the intrinsic endpoint at \(\Theta=\pi\) and shows that the corresponding physical azimuth reaches \(\phi(\pi)=\pi+\alpha\). The first description therefore treats bending as a displacement of an orbital root, whereas the second treats it as excess azimuth accumulated during a fixed phase advance. This equivalence motivates the subsequent calculation in which the interval in \(\Theta\) remains unchanged at every perturbative order.

An exact relation inherited from the radial first integral provides an independent consistency check on the phase equations. Substitution of Eq. \eqref{3.2} into Eq. \eqref{2.13} gives
\begin{equation}
A^{2}\cos^{2}\Theta
=
1
-
A^{2}\sin^{2}\Theta
+
2\epsilon
A^{3}\sin^{3}\Theta.
\end{equation}
Using \(\sin^{2}\Theta+\cos^{2}\Theta=1\), we obtain
\begin{equation}
A^{2}
=
1
+
2\epsilon
A^{3}\sin^{3}\Theta.
\label{3.43}
\end{equation}
This identity shows that the Schwarzschild amplitude is not an independent arbitrary function once the phase is specified. The nonlinear orbit dynamics impose an exact relation between \(A\), \(\Theta\), and \(\epsilon\).

We can also check that Eq. \eqref{3.43} agrees with the differential amplitude equation. Differentiating Eq. \eqref{3.43} with respect to \(\phi\), using Eq. \eqref{3.14}, and simplifying reproduces Eq. \eqref{3.12}. The differential and first-integral formulations therefore describe the same phase-plane flow.

The resulting structure separates the scattering problem into two logically distinct parts. The phase variable always travels from \(0\) to \(\pi\), while the amplitude determines how strongly the physical azimuth departs from this intrinsic phase advance. The exact bending angle then follows from Eq. \eqref{3.35} without locating a perturbed outgoing zero. This fixed-endpoint formulation supplies the mathematical basis for treating weak Schwarzschild deflection as a phase-advance problem.

\section{Algebraic Reduction of the Schwarzschild Amplitude}
\label{sec4}
We now exploit a special property of the Schwarzschild phase equations that substantially simplifies the scattering problem. The exact phase formulation derived in Sec. \ref{sec3} gives
\begin{equation}
\frac{dA}{d\phi}
=
3\epsilon
A^{2}
\sin^{2}\Theta
\cos\Theta
\label{4.1}
\end{equation}
and
\begin{equation}
\frac{d\Theta}{d\phi}
=
1
-
3\epsilon
A\sin^{3}\Theta.
\label{4.2}
\end{equation}
The amplitude \(A\) and phase \(\Theta\) remain coupled in these equations. For Schwarzschild spacetime, however, their coupling contains a repeated combination of the phase that allows us to remove the differential amplitude evolution entirely.

Since the phase increases monotonically along every returning trajectory considered here, we may regard \(A\) as a function of \(\Theta\). Dividing Eq. \eqref{4.1} by Eq. \eqref{4.2} gives
\begin{equation}
\frac{dA}{d\Theta}
=
\frac{
3\epsilon
A^{2}
\sin^{2}\Theta
\cos\Theta
}
{
1
-
3\epsilon
A\sin^{3}\Theta
}.
\label{4.3}
\end{equation}
The dependence on \(\Theta\) suggests that we introduce the composite variable
\begin{equation}
q
=
\epsilon\sin^{3}\Theta.
\label{4.4}
\end{equation}
Its derivative is
\begin{equation}
\frac{dq}{d\Theta}
=
3\epsilon
\sin^{2}\Theta
\cos\Theta.
\label{4.5}
\end{equation}
Combining Eqs. \eqref{4.3} and \eqref{4.5} gives
\begin{equation}
\frac{dA}{dq}
=
\frac{
A^{2}
}
{
1-3qA
}.
\label{4.6}
\end{equation}

The reduction in Eq. \eqref{4.6} is important because the explicit phase dependence has disappeared. We now have an autonomous first-order equation involving only \(A\) and \(q\). More importantly, the equation becomes linear after we interchange the roles of the dependent and independent variables. Inverting Eq. \eqref{4.6} gives
\begin{equation}
\frac{dq}{dA}
=
\frac{1}{A^{2}}
-
\frac{3q}{A}.
\end{equation}
We therefore obtain
\begin{equation}
\frac{dq}{dA}
+
\frac{3}{A}q
=
\frac{1}{A^{2}}.
\end{equation}
Multiplication by the integrating factor \(A^{3}\) yields
\begin{equation}
\frac{d}{dA}
\left(
A^{3}q
\right)
=
A.
\end{equation}
Integration gives
\begin{equation}
A^{3}q
=
\frac{A^{2}}{2}
+
C.
\label{4.10}
\end{equation}

At either asymptotic endpoint we have \(q=0\) because \(\sin\Theta=0\). We also established in Sec. \ref{sec3} that \(A=1\) at the asymptotes. Evaluating Eq. \eqref{4.10} there fixes the integration constant as \(C = -1/2\).
Hence the exact amplitude satisfies
\begin{equation}
2qA^{3}
-
A^{2}
+
1
=
0.
\label{4.12}
\end{equation}

The result in Eq. \eqref{4.12} converts the amplitude problem from a differential problem into an algebraic one. Once \(\Theta\) is specified, Eq. \eqref{4.4} fixes \(q\), and Eq. \eqref{4.12} fixes the physical value of \(A\). We therefore need not integrate Eq. \eqref{4.1} along the null trajectory.

The same algebraic relation follows directly from the first-integral identity \eqref{3.43}: using Eq. \eqref{4.4} in that identity gives Eq. \eqref{4.12}. The differential reduction and the radial first integral therefore lead to the same result.

A reciprocal amplitude makes the structure still simpler. We define \(y=1/A\). Multiplying Eq. \eqref{4.12} by \(y^{3}\) gives
\begin{equation}
2q
=
y-y^{3}.
\label{4.16}
\end{equation}
We have thus reduced the exact Schwarzschild amplitude to the inverse of a cubic map.

The asymptotic condition \(A=1\) becomes
\begin{equation}
y(0)=1.
\label{4.17}
\end{equation}
Along the incoming branch, \(q\) increases from zero to its maximum value at closest approach. Since \(q=\epsilon\sin^{3}\Theta\), we have \(0 \leq q \leq \epsilon\), with \(q=\epsilon\) at \(\Theta=\pi/2\). The variable \(q\) then decreases symmetrically back to zero on the outgoing branch.

For a returning Schwarzschild trajectory we established \(0<\epsilon<1/(3\sqrt{3})\). The physical solution of Eq. \eqref{4.16} that connects continuously to \(y=1\) at \(q=0\) remains within \(1/\sqrt{3}<y\leq1\). Consequently, \(1\leq A<\sqrt{3}\). The amplitude grows from unity as the photon approaches the turning point and decreases back to unity after closest approach.

We may solve the cubic in Eq. \eqref{4.16} exactly on this physical branch. Writing
\begin{equation}
y
=
\frac{2}{\sqrt{3}}
\cos\chi,
\end{equation}
and using
\begin{equation}
4\cos^{3}\chi
-
3\cos\chi
=
\cos 3\chi,
\end{equation}
we find
\begin{equation}
\cos 3\chi
=
-3\sqrt{3}\,q.
\end{equation}
The branch satisfying Eq. \eqref{4.17} is therefore
\begin{equation}
\chi(q)
=
\frac{1}{3}
\arccos
\left(
-3\sqrt{3}\,q
\right),
\end{equation}
which gives
\begin{equation}
y(q)
=
\frac{2}{\sqrt{3}}
\cos
\left[
\frac{1}{3}
\arccos
\left(
-3\sqrt{3}\,q
\right)
\right].
\label{4.29}
\end{equation}
The corresponding amplitude is
\begin{equation}
A(q)
=
\frac{
\sqrt{3}
}
{
2
\cos
\left[
\frac{1}{3}
\arccos
\left(
-3\sqrt{3}\,q
\right)
\right]
}.
\end{equation}
Thus the Schwarzschild phase-plane amplitude can be written exactly as an elementary algebraic function of \(q=\epsilon\sin^{3}\Theta\).

The cubic relation also simplifies the local phase rate. Since \(A=1/y\), the factor appearing in the phase equation becomes
\begin{equation}
1-3qA
=
1-\frac{3q}{y}.
\end{equation}
Using Eq. \eqref{4.16} gives
\begin{equation}
1-\frac{3q}{y}
=
1
-
\frac{3}{2}
\left(
1-y^{2}
\right),
\end{equation}
and hence
\begin{equation}
1-3qA
=
\frac{
3y^{2}-1
}{2}.
\end{equation}
The physical phase rate in Eq. \eqref{4.2} therefore reduces to
\begin{equation}
\frac{d\Theta}{d\phi}
=
\frac{
3y^{2}-1
}{2}.
\label{4.35}
\end{equation}
Its reciprocal is
\begin{equation}
\frac{d\phi}{d\Theta}
=
\frac{2}
{
3y^{2}-1
}.
\label{4.36}
\end{equation}

We can express the same quantity entirely through the cubic map. Differentiating Eq. \eqref{4.16} with respect to \(q\) gives
\begin{equation}
2
=
\left(
1-3y^{2}
\right)
\frac{dy}{dq}.
\end{equation}
Therefore
\begin{equation}
\frac{dy}{dq}
=
-\frac{2}
{
3y^{2}-1
}.
\label{4.38}
\end{equation}
Comparison of Eqs. \eqref{4.36} and \eqref{4.38} yields the exact identity
\begin{equation}
\frac{d\phi}{d\Theta}
=
-\frac{dy}{dq}.
\label{4.39}
\end{equation}
This relation provides the most compact form of the Schwarzschild algebraic reduction. The physical azimuthal rate with respect to the intrinsic phase equals the negative derivative of the reciprocal amplitude with respect to the composite variable \(q\).

We may now rewrite the exact phase expression for the bending angle obtained in Sec. \ref{sec3}. Substitution of Eq. \eqref{4.39} gives
\begin{equation}
\alpha
=
\int_{0}^{\pi}
\left[
-
\frac{dy}{dq}
\bigg|_{
q=\epsilon\sin^{3}\Theta
}
-1
\right]
d\Theta.
\end{equation}
The symmetry
\begin{equation}
\sin(\pi-\Theta)
=
\sin\Theta
\end{equation}
implies that the integrand is symmetric about closest approach. We may therefore write
\begin{equation}
\alpha
=
2
\int_{0}^{\pi/2}
\left[
-
\frac{dy}{dq}
\bigg|_{
q=\epsilon\sin^{3}\Theta
}
-1
\right]
d\Theta.
\end{equation}

An exact algebraic phase factor will be useful below. We define
\begin{equation}
\mathcal{F}(q)
=
-\frac{dy}{dq}.
\end{equation}
Using Eq. \eqref{4.38}, we find
\begin{equation}
\mathcal{F}(q)
=
\frac{2}
{
3y^{2}(q)-1
}.
\label{4.44}
\end{equation}
With the explicit physical branch in Eq. \eqref{4.29}, this becomes
\begin{equation}
\mathcal{F}(q)
=
\frac{2}
{
4
\cos^{2}
\left[
\frac{1}{3}
\arccos
\left(
-3\sqrt{3}\,q
\right)
\right]
-1
}.
\end{equation}
The exact Schwarzschild deflection angle can consequently be written as
\begin{equation}
\alpha
=
\int_{0}^{\pi}
\left[
\mathcal{F}
\left(
\epsilon\sin^{3}\Theta
\right)
-1
\right]
d\Theta.
\label{4.46}
\end{equation}

Equation \eqref{4.46} isolates the entire nonlinear Schwarzschild dynamics in the single algebraic function \(\mathcal{F}(q)\). The remaining dependence on the orbital phase enters only through the elementary combination \(\sin^{3}\Theta\). We therefore replace the original nonlinear second-order orbit problem by a local cubic inversion followed by a fixed-interval phase integral.

\begin{figure}[t]
\centering
\includegraphics[width=\columnwidth]{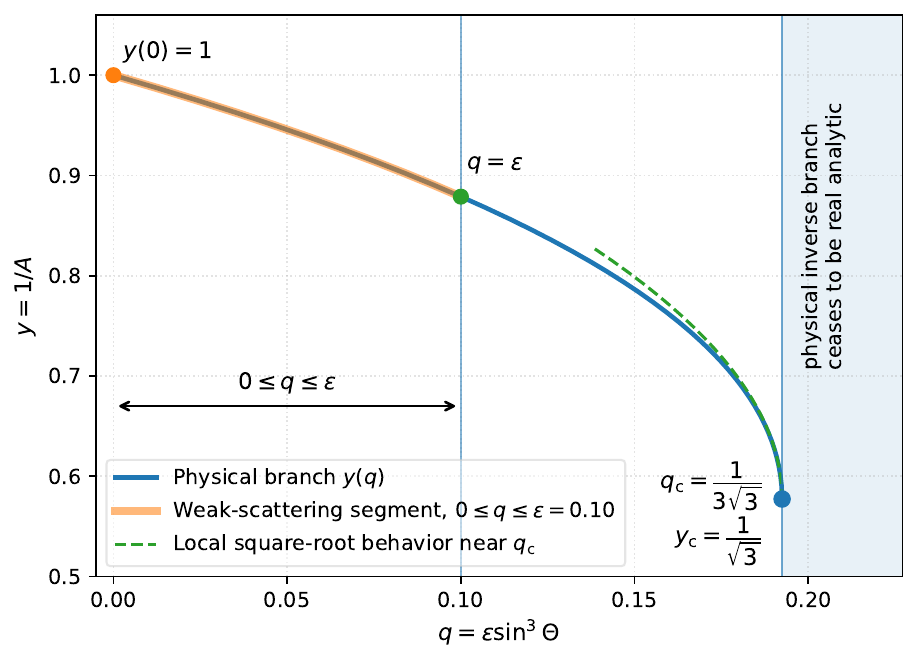}
\caption{Physical inverse branch of \(2q=y-y^{3}\), showing the weak-scattering interval and the critical square-root branch point at \(q_{\rm c}=1/(3\sqrt{3})\).}
\label{fig3}
\end{figure}
The algebraic branch clarifies both the weak-scattering interval and the origin of the limiting singularity (see Fig. \ref{fig3}). Starting from the asymptotic point \(y(0)=1\), the physical branch decreases smoothly as \(q\) increases and remains regular throughout \(0\leq q\leq\epsilon\) for every returning orbit with \(\epsilon<1/(3\sqrt{3})\). The inverse relation loses analyticity at \(q_{\rm c}=1/(3\sqrt{3})\) and \(y_{\rm c}=1/\sqrt{3}\), where \(dq/dy\) vanishes and \(dy/dq\) diverges. The highlighted weak interval shows that an ordinary weak ray samples only a proper subset of the analytic branch, while the local square-root curve illustrates how the physical solution approaches its branch point. This structure provides the mathematical reason that the perturbative expansion about \(q=0\) possesses a finite convergence radius.

For the weak-field expansion, it is convenient to measure the departure of the reciprocal amplitude from its asymptotic value. We define \(\delta =  1-y\). The physical weak-scattering branch satisfies \(\delta\rightarrow0\) as \(q\rightarrow0\). Substituting \(y=1-\delta\) into Eq. \eqref{4.16} gives
\begin{equation}
2q
=
(1-\delta)
-
(1-\delta)^{3}.
\end{equation}
Expanding the cubic exactly gives
\begin{equation}
2q
=
2\delta
-
3\delta^{2}
+
\delta^{3}.
\end{equation}
Hence
\begin{equation}
q
=
\delta
-
\frac{3}{2}\delta^{2}
+
\frac{1}{2}\delta^{3}.
\end{equation}
Equivalently, we may write
\begin{equation}
q
=
\delta
\left(
1
-
\frac{3}{2}\delta
+
\frac{1}{2}\delta^{2}
\right).
\label{4.52}
\end{equation}

The weak-deflection calculation has now reached a purely algebraic form. We need only invert the cubic map in Eq. \eqref{4.52} around \(\delta=0\), use the resulting series to expand the exact phase factor \(\mathcal{F}(q)\), and integrate powers of \(\sin\Theta\) over the fixed interval from \(0\) to \(\pi\). No perturbative solution of the null orbit remains necessary.

\section{All-Order Weak-Deflection Expansion}
\label{sec5}
We now convert the exact algebraic phase formulation into an explicit weak-field expansion valid to arbitrary perturbative order. The reduction obtained in Sec. \ref{sec4} gives
\begin{equation}
q
=
\delta
\left(
1-\delta
\right)
\left(
1-\frac{\delta}{2}
\right).
\label{5.1}
\end{equation}
where
\begin{equation}
\delta
=
1-y
=
1-\frac{1}{A}.
\end{equation}
The physical branch satisfies \(\delta=0\) when \(q=0\). Since the derivative of the right-hand side of Eq. \eqref{5.1} with respect to \(\delta\) equals unity at \(\delta=0\), the inverse function exists analytically in a neighborhood of the weak-field point.

We write Eq. \eqref{5.1} in the form \(\delta = q\,\Psi(\delta),\) where
\begin{equation}
\Psi(\delta)
=
\frac{1}
{
(1-\delta)
\left(
1-\frac{\delta}{2}
\right)
}.
\label{5.4}
\end{equation}
This form allows us to apply the Lagrange inversion formula directly. We introduce the expansion
\begin{equation}
\delta(q)
=
\sum_{n=1}^{\infty}
d_{n}q^{n}.
\label{5.5}
\end{equation}
Lagrange inversion gives
\begin{equation}
d_{n}
=
\frac{1}{n}
[t^{\,n-1}]
\Psi^{n}(t),
\label{5.6}
\end{equation}
where \([t^{m}]F(t)\) denotes the coefficient multiplying \(t^{m}\) in the Taylor expansion of \(F(t)\).

Using Eq. \eqref{5.4}, we obtain
\begin{equation}
d_{n}
=
\frac{1}{n}
[t^{\,n-1}]
(1-t)^{-n}
\left(
1-\frac{t}{2}
\right)^{-n}.
\end{equation}
The first several coefficients are
\begin{align}
d_{1}=1,
\qquad
d_{2}=\frac{3}{2},
\qquad
d_{3}=4,
\qquad
d_{4}=\frac{105}{8},
\end{align}
\begin{align}
d_{5}=48,
\quad
d_{6}=\frac{3003}{16},
\quad
d_{7}=768,
\quad
d_{8}=\frac{415701}{128}.
\end{align}
Hence the reciprocal-amplitude displacement takes the form
\begin{align}
\delta(q)
={}&
q
+
\frac{3}{2}q^{2}
+
4q^{3}
+
\frac{105}{8}q^{4}
+
48q^{5}
+
\frac{3003}{16}q^{6} \nn
&+
768q^{7}
+
\frac{415701}{128}q^{8}
+
\mathcal{O}(q^{9}).
\end{align}

The bending angle does not require \(\delta(q)\) itself. Since \(y=1-\delta\), the phase factor defined in Sec. \ref{sec4} satisfies the particularly simple relation
\begin{equation}
\mathcal{F}(q)
=
\frac{d\delta}{dq}.
\end{equation}
We therefore write
\begin{equation}
\mathcal{F}(q)
=
1
+
\sum_{n=1}^{\infty}
f_{n}q^{n}.
\label{5.13}
\end{equation}
Differentiation of Eq. \eqref{5.5} gives
\begin{equation}
f_{n}
=
(n+1)d_{n+1}.
\label{5.14}
\end{equation}
Combining Eqs. \eqref{5.6} and \eqref{5.14}, we find
\begin{equation}
f_{n}
=
[t^{n}]
\Psi^{n+1}(t).
\end{equation}
Equation \eqref{5.4} then gives
\begin{equation}
f_{n}
=
[t^{n}]
(1-t)^{-(n+1)}
\left(
1-\frac{t}{2}
\right)^{-(n+1)}.
\label{5.16}
\end{equation}

We can convert the coefficient extraction in Eq. \eqref{5.16} into a finite combinatorial expression. The two factors possess the expansions
\begin{equation}
(1-t)^{-(n+1)}
=
\sum_{r=0}^{\infty}
\binom{n+r}{n}
t^{r}
\end{equation}
and
\begin{equation}
\left(
1-\frac{t}{2}
\right)^{-(n+1)}
=
\sum_{s=0}^{\infty}
2^{-s}
\binom{n+s}{n}
t^{s}.
\end{equation}
The coefficient of \(t^{n}\) follows by setting \(r=n-s\). We obtain
\begin{equation}
f_{n}
=
\sum_{s=0}^{n}
2^{-s}
\binom{2n-s}{n}
\binom{n+s}{n}.
\label{5.19}
\end{equation}
This finite sum provides an explicit expression for the local Schwarzschild phase coefficient at arbitrary order.

The first coefficients read
\begin{align}
f_{0}=1,
\qquad
f_{1}=3,
\qquad
f_{2}=12,
\qquad
f_{3}=\frac{105}{2},
\end{align}
\begin{align}
f_{4}=240,
\qquad
f_{5}=\frac{9009}{8},
\nn
f_{6}=5376,
\qquad
f_{7}=\frac{415701}{16}.
\end{align}
Thus the exact phase factor has the weak-field expansion
\begin{align}
&\mathcal{F}(q)
=
1
+
3q
+
12q^{2}
+
\frac{105}{2}q^{3}
+
240q^{4} \nn
&+
\frac{9009}{8}q^{5}
+
5376q^{6}
+
\frac{415701}{16}q^{7}
+
\mathcal{O}(q^{8}).
\end{align}

We can obtain another compact generating relation without performing the inverse series explicitly. Solving Eq. \eqref{4.44} for \(y^{2}\) yields
\begin{equation}
y^{2}
=
\frac{
\mathcal{F}+2
}
{
3\mathcal{F}
}.
\label{5.25}
\end{equation}
Squaring Eq. \eqref{4.16} gives
\begin{equation}
4q^{2}
=
y^{2}
\left(
1-y^{2}
\right)^{2}.
\label{5.26}
\end{equation}
Substitution of Eq. \eqref{5.25} into Eq. \eqref{5.26} leads to
\begin{equation}
27q^{2}\mathcal{F}^{3}
=
\left(
\mathcal{F}+2
\right)
\left(
\mathcal{F}-1
\right)^{2}.
\end{equation}
After expansion, we obtain the algebraic generating equation
\begin{equation}
\left(
1-27q^{2}
\right)
\mathcal{F}^{3}
-
3\mathcal{F}
+
2
=
0.
\label{5.28}
\end{equation}
The physical solution is the branch satisfying
\begin{align}
\mathcal{F}(0)=1,
\qquad
\left.
\frac{d\mathcal{F}}{dq}
\right|_{q=0}
=
3.
\end{align}
Equations \eqref{5.19} and \eqref{5.28} give two complementary all-order descriptions. The first provides each coefficient directly through a finite sum. The second packages the complete coefficient sequence into one algebraic relation.

We now insert the local phase expansion into the exact bending expression \eqref{4.46}. For sufficiently small \(\epsilon\), substitution of Eq. \eqref{5.13} term by term gives
\begin{equation}
\alpha
=
\sum_{n=1}^{\infty}
f_{n}
\epsilon^{n}
\int_{0}^{\pi}
\sin^{3n}\Theta\,d\Theta.
\label{5.31}
\end{equation}
The nonlinear Schwarzschild dynamics and the phase integration have therefore separated completely.

We define the universal trigonometric moment
\begin{equation}
I_{m}
=
\int_{0}^{\pi}
\sin^{m}\Theta\,d\Theta.
\end{equation}
Reflection symmetry about \(\Theta=\pi/2\) gives
\begin{equation}
I_{m}
=
2
\int_{0}^{\pi/2}
\sin^{m}\Theta\,d\Theta.
\end{equation}
Using the Euler beta integral, we find
\begin{equation}
I_{m}
=
B
\left(
\frac{m+1}{2},
\frac{1}{2}
\right)
\end{equation}
and therefore
\begin{equation}
I_{m}
=
\frac{
\sqrt{\pi}\,
\Gamma
\left(
\frac{m+1}{2}
\right)
}
{
\Gamma
\left(
\frac{m+2}{2}
\right)
}.
\label{5.35}
\end{equation}

Substitution of Eq. \eqref{5.35} into Eq. \eqref{5.31} produces the all-order Schwarzschild weak-deflection series
\begin{equation}
\alpha(\epsilon)
=
\sum_{n=1}^{\infty}
\mathcal{A}_{n}
\epsilon^{n},
\label{5.36}
\end{equation}
where
\begin{equation}
\mathcal{A}_{n}
=
f_{n}
\frac{
\sqrt{\pi}\,
\Gamma
\left(
\frac{3n+1}{2}
\right)
}
{
\Gamma
\left(
\frac{3n+2}{2}
\right)
}.
\label{5.37}
\end{equation}
Using the finite expression in Eq. \eqref{5.19}, we arrive at
\begin{equation}
\boxed{
\mathcal{A}_{n}
=
\frac{
\sqrt{\pi}\,
\Gamma
\left(
\frac{3n+1}{2}
\right)
}
{
\Gamma
\left(
\frac{3n+2}{2}
\right)
}
\sum_{s=0}^{n}
2^{-s}
\binom{2n-s}{n}
\binom{n+s}{n}
}.
\label{5.38}
\end{equation}
Equation \eqref{5.38} gives the Schwarzschild weak-deflection coefficient at arbitrary order directly from the order \(n\). We do not need to solve a hierarchy of orbit equations before determining \(\mathcal{A}_{n}\).

The factorization in Eq. \eqref{5.37} also explains the alternating analytic structure of the Schwarzschild weak series. For even perturbative order, we set
\begin{equation}
n=2j.
\end{equation}
The relevant trigonometric moment becomes
\begin{equation}
I_{6j}
=
\pi
\frac{
(6j)!
}
{
2^{6j}
\left[
(3j)!
\right]^{2}
}.
\end{equation}
Hence
\begin{equation}
\mathcal{A}_{2j}
=
\pi f_{2j}
\frac{
(6j)!
}
{
2^{6j}
\left[
(3j)!
\right]^{2}
}.
\label{5.41}
\end{equation}

For odd perturbative order, we write
\begin{equation}
n=2j+1.
\end{equation}
We then obtain
\begin{equation}
I_{6j+3}
=
\frac{
2^{6j+3}
\left[
(3j+1)!
\right]^{2}
}
{
(6j+3)!
}.
\end{equation}
The corresponding coefficient is
\begin{equation}
\mathcal{A}_{2j+1}
=
f_{2j+1}
\frac{
2^{6j+3}
\left[
(3j+1)!
\right]^{2}
}
{
(6j+3)!
}.
\label{5.44}
\end{equation}
Since every \(f_{n}\) in Eq. \eqref{5.19} is rational, Eqs. \eqref{5.41} and \eqref{5.44} show directly why even powers of \(M/b\) carry a factor of \(\pi\), while odd powers remain rational. The distinction arises entirely from the universal phase moments rather than from separate dynamical calculations at odd and even order.

At first order, Eq. \eqref{5.19} gives \(f_{1}=3\). Since \(I_{3} = 4/3\),  we obtain \(\mathcal{A}_{1} = 4\). At second order, \(f_{2} = 12\), and \(I_{6} = 5\pi/16\), which gives \(\mathcal{A}_{2} = 15\pi/4\). At third order, \(f_{3} = 105/2\), \(I_{9} = 256/315\) so that \(\mathcal{A}_{3} = 128/3\). At fourth order, \(f_{4} = 240,\) \(I_{12} = 231/pi/1024\), which gives \(\mathcal{A}_{4}=3465\pi/64\).

Continuing the same algebraic procedure gives
\begin{align}
&\alpha
=
4\epsilon
+
\frac{15\pi}{4}\epsilon^{2}
+
\frac{128}{3}\epsilon^{3}
+
\frac{3465\pi}{64}\epsilon^{4}
+
\frac{3584}{5}\epsilon^{5} \nn 
&+
\frac{255255\pi}{256}\epsilon^{6}
+
\frac{98304}{7}\epsilon^{7}
+
\frac{334639305\pi}{16384}\epsilon^{8}
+
\mathcal{O}(\epsilon^{9}).
\end{align}
Restoring \(\epsilon=M/b\), we obtain
\begin{align}
&\alpha(b)
=
\frac{4M}{b}
+
\frac{15\pi}{4}
\left(
\frac{M}{b}
\right)^{2}
+
\frac{128}{3}
\left(
\frac{M}{b}
\right)^{3} 
\nn
&+
\frac{3465\pi}{64}
\left(
\frac{M}{b}
\right)^{4} +
\frac{3584}{5}
\left(
\frac{M}{b}
\right)^{5}
\nn &+
\frac{255255\pi}{256}
\left(
\frac{M}{b}
\right)^{6}
+
\frac{98304}{7}
\left(
\frac{M}{b}
\right)^{7}
\\
&+
\frac{334639305\pi}{16384}
\left(
\frac{M}{b}
\right)^{8}
+
\mathcal{O}
\left[
\left(
\frac{M}{b}
\right)^{9}
\right].
\end{align}

The structure of Eq. \eqref{5.38} constitutes the main mathematical gain of the phase-plane method. At perturbative order \(n\), the local Schwarzschild dynamics contribute only the algebraic coefficient \(f_{n}\), while the global passage from the incoming to the outgoing asymptote contributes only the universal phase moment \(I_{3n}\). We therefore factor every weak-deflection coefficient into a local nonlinear part and a fixed-interval geometrical part.

We can summarize the result in a compact statement. For the Schwarzschild null-scattering problem with invariant weak parameter \(\epsilon\), the phase-plane formulation gives
\begin{equation}
\boxed{
\begin{aligned}
\alpha(\epsilon)
={}&
\sum_{n=1}^{\infty}
\left[
\sum_{s=0}^{n}
2^{-s}
\binom{2n-s}{n}
\binom{n+s}{n}
\right]
\\
&\times
\frac{
\sqrt{\pi}\,
\Gamma
\left(
\frac{3n+1}{2}
\right)
}
{
\Gamma
\left(
\frac{3n+2}{2}
\right)
}
\epsilon^{n}
\end{aligned}
}.
\label{5.58}
\end{equation}
Within the weak-field branch, Eq. \eqref{5.58} replaces the usual order-by-order solution of the nonlinear orbit equation by an explicit coefficient rule. The entire perturbative calculation follows from inversion of one cubic algebraic map and evaluation of one family of elementary phase moments.

\section{Consistency Checks and Comparison with Standard Schwarzschild Bending}
\label{sec6}
We now test the phase-plane formulation against independent descriptions of Schwarzschild null scattering. Agreement of the first few weak-field coefficients alone would not provide a sufficiently strong check because those coefficients arise from the same underlying geodesic dynamics. We therefore compare the phase result with the exact radial scattering integral, reproduce the first two coefficients through the conventional orbit perturbation, examine the numerical behavior of successive truncations, and verify that the phase formulation identifies the same critical scattering boundary as the original radial equation.

A direct comparison begins with the exact first integral \eqref{2.13}. Let \(z_{\mathrm{m}}\) denote the positive turning-point value reached by a returning photon. It satisfies
\begin{equation}
1-z_{\mathrm{m}}^{2}
+
2\epsilon z_{\mathrm{m}}^{3}
=
0.
\label{6.2}
\end{equation}
Along the incoming branch we have \(z'>0\), while along the outgoing branch we have \(z'<0\). Reflection symmetry about closest approach therefore gives the exact total azimuthal change
\begin{equation}
\Delta\phi_{\mathrm{rad}}
=
2
\int_{0}^{z_{\mathrm{m}}}
\frac{
dz
}{
\sqrt{
1-z^{2}
+
2\epsilon z^{3}
}
}.
\end{equation}
The corresponding bending angle is
\begin{equation}
\alpha_{\mathrm{rad}}
=
2
\int_{0}^{z_{\mathrm{m}}}
\frac{
dz
}{
\sqrt{
1-z^{2}
+
2\epsilon z^{3}
}
}
-\pi.
\label{6.4}
\end{equation}

The upper endpoint in Eq. \eqref{6.4} is an integrable square-root singularity. For numerical evaluation and perturbative comparison, we remove this endpoint behavior by writing \(z=z_{\mathrm{m}}\sin\chi,\) and \(0\leq\chi\leq \pi/2\). Using the turning-point condition in Eq. \eqref{6.2}, the polynomial under the square root factorizes as
\begin{align}
&1
-
z_{\mathrm{m}}^{2}\sin^{2}\chi
+
2\epsilon z_{\mathrm{m}}^{3}\sin^{3}\chi
=
z_{\mathrm{m}}^{2}
\left(
1-\sin\chi
\right)
\\
&\times
\left[
1+\sin\chi
-
2\epsilon z_{\mathrm{m}}
\left(
1+\sin\chi+\sin^{2}\chi
\right)
\right].
\end{align}
Since
\begin{equation}
\cos^{2}\chi
=
\left(
1-\sin\chi
\right)
\left(
1+\sin\chi
\right),
\end{equation}
Eq. \eqref{6.4} becomes
\begin{align}
&\alpha_{\mathrm{rad}}
=
2
\int_{0}^{\pi/2} d\chi \nn &
\times \left[
\frac{
1+\sin\chi
}{
1+\sin\chi
-
2\epsilon z_{\mathrm{m}}
\left(
1+\sin\chi+\sin^{2}\chi
\right)
}
\right]^{1/2}
-\pi.
\label{6.8}
\end{align}

The integrand remains finite at \(\chi=\pi/2\) for every returning orbit with \(b>3\sqrt{3}M\).

For weak scattering, the turning point itself admits an expansion in \(\epsilon\). We write
\begin{equation}
z_{\mathrm{m}}
=
1
+
c_{1}\epsilon
+
c_{2}\epsilon^{2}
+
c_{3}\epsilon^{3}
+
c_{4}\epsilon^{4}
+
\mathcal{O}(\epsilon^{5}).
\end{equation}
Substitution into Eq. \eqref{6.2} and comparison of equal powers of \(\epsilon\) gives
\begin{align}
c_{1}=1,
\qquad
c_{2}=\frac{5}{2},
\qquad
c_{3}=8,
\qquad
c_{4}=\frac{231}{8}.
\end{align}
Hence
\begin{equation}
z_{\mathrm{m}}
=
1
+
\epsilon
+
\frac{5}{2}\epsilon^{2}
+
8\epsilon^{3}
+
\frac{231}{8}\epsilon^{4}
+
\mathcal{O}(\epsilon^{5}).
\label{6.11}
\end{equation}

Expanding the regular expression in Eq. \eqref{6.8} with Eq. \eqref{6.11}, followed by termwise integration over \(\chi\), gives
\begin{equation}
\alpha_{\mathrm{rad}}
=
4\epsilon
+
\frac{15\pi}{4}\epsilon^{2}
+
\frac{128}{3}\epsilon^{3}
+
\frac{3465\pi}{64}\epsilon^{4}
+
\mathcal{O}(\epsilon^{5}).
\end{equation}
The coefficients agree exactly with those obtained from the phase coefficient formula in Sec. \ref{sec5}. This comparison is important because Eq. \eqref{6.8} follows directly from the radial first integral and contains neither the phase variable \(\Theta\) nor the algebraic phase factor \(\mathcal{F}\).

A simpler check isolates the leading Einstein term directly within the phase representation. Since \(A=1+\mathcal{O}(\epsilon)\), expansion of Eq. \eqref{3.35} and use of \(\int_{0}^{\pi}\sin^{3}\Theta\,d\Theta=4/3\) give
\begin{align}
\alpha
&=
3\epsilon
\int_{0}^{\pi}
\sin^{3}\Theta\,d\Theta
+
\mathcal{O}(\epsilon^{2})
\nn
&=
4\epsilon
+
\mathcal{O}(\epsilon^{2})
=
\frac{4M}{b}
+
\mathcal{O}
\left[
\left(
\frac{M}{b}
\right)^{2}
\right].
\end{align}
The leading weak bending thus follows from one fixed phase moment.

For a more independent perturbative check, we return temporarily to the conventional Schwarzschild orbit equation and solve it order by order. We write
\begin{equation}
z(\phi)
=
Z_{0}(\phi)
+
\epsilon Z_{1}(\phi)
+
\epsilon^{2}Z_{2}(\phi)
+
\mathcal{O}(\epsilon^{3}).
\end{equation}
The asymptotic initial conditions imply
\begin{align}
Z_{0}(0)=0,
\qquad
Z_{0}'(0)=1,
\label{6.20}
\end{align}
and
\begin{align}
Z_{n}(0)=0,
\qquad
Z_{n}'(0)=0
\label{6.21}
\end{align}
for \(n \geq 1\)

At zeroth order, the orbit equation gives
\begin{equation}
Z_{0}''+Z_{0}=0.
\end{equation}
The solution satisfying Eq. \eqref{6.20} is
\begin{equation}
Z_{0}
=
\sin\phi.
\label{6.23}
\end{equation}

At first order, we obtain
\begin{equation}
Z_{1}''+Z_{1}
=
3Z_{0}^{2}.
\end{equation}
Using Eq. \eqref{6.23}, the source becomes
\begin{equation}
3Z_{0}^{2}
=
\frac{3}{2}
\left(
1-\cos2\phi
\right).
\end{equation}
The solution satisfying Eq. \eqref{6.21} is
\begin{equation}
Z_{1}
=
\frac{3}{2}
+
\frac{1}{2}\cos2\phi
-
2\cos\phi.
\end{equation}
An equivalent and more compact form is
\begin{equation}
Z_{1}
=
\left(
1-\cos\phi
\right)^{2}.
\label{6.27}
\end{equation}

At second order, the orbit equation gives
\begin{equation}
Z_{2}''+Z_{2}
=
6Z_{0}Z_{1}.
\end{equation}
Using Eqs. \eqref{6.23} and \eqref{6.27}, we obtain
\begin{equation}
Z_{2}''+Z_{2}
=
6
\sin\phi
\left(
1-\cos\phi
\right)^{2}.
\end{equation}
The unique solution satisfying the homogeneous initial conditions in Eq. \eqref{6.21} is
\begin{equation}
Z_{2}
=
-\frac{15}{4}\phi\cos\phi
+
\frac{3}{4}\sin^{3}\phi
-
\frac{1}{4}\sin\phi
+
2\sin2\phi.
\label{6.30}
\end{equation}

We now locate the outgoing asymptote in the conventional manner. We write
\begin{equation}
\alpha
=
a_{1}\epsilon
+
a_{2}\epsilon^{2}
+
\mathcal{O}(\epsilon^{3})
\end{equation}
and impose the outgoing condition in Eq. \eqref{2.26}.
The zeroth-order contribution gives
\begin{equation}
Z_{0}(\pi+\alpha)
=
-\alpha
+
\mathcal{O}(\alpha^{3}),
\end{equation}
while Eq. \eqref{6.27} gives
\begin{align}
Z_{1}(\pi)=4,
\qquad
Z_{1}'(\pi)=0.
\end{align}
From Eq. \eqref{6.30}, we obtain
\begin{equation}
Z_{2}(\pi)
=
\frac{15\pi}{4}.
\end{equation}
Substitution into Eq. \eqref{2.26} therefore gives
\begin{equation}
0
=
\left(
-a_{1}+4
\right)\epsilon
+
\left(
-a_{2}
+
\frac{15\pi}{4}
\right)
\epsilon^{2}
+
\mathcal{O}(\epsilon^{3}).
\end{equation}
Equality at each perturbative order requires
\begin{align}
a_{1}=4,
\qquad
a_{2}=\frac{15\pi}{4}.
\end{align}
Thus the conventional displaced-root calculation yields
\begin{equation}
\alpha
=
4\epsilon
+
\frac{15\pi}{4}\epsilon^{2}
+
\mathcal{O}(\epsilon^{3}),
\end{equation}
in exact agreement with the phase-plane result.

The comparison also reveals the difference in mathematical organization. In the conventional calculation, the second-order coefficient requires the explicit functions \(Z_{1}(\phi)\) and \(Z_{2}(\phi)\), followed by an expansion of the displaced root. The phase formulation instead obtains the same coefficient from the local algebraic coefficient \(f_{2}=12\) and the fixed phase moment
\begin{equation}
I_{6}
=
\int_{0}^{\pi}
\sin^{6}\Theta\,d\Theta
=
\frac{5\pi}{16}.
\end{equation}
Their product gives
\begin{equation}
f_{2}I_{6}
=
\frac{15\pi}{4}.
\end{equation}
The two calculations therefore agree while acting on different mathematical objects.

A numerical comparison provides another check that does not rely on matching a finite number of symbolic coefficients. We define the truncation through order \(N\) by
\begin{equation}
\alpha^{(N)}(\epsilon)
=
\sum_{n=1}^{N}
\mathcal{A}_{n}\epsilon^{n}
\end{equation}
and measure its relative deviation from the exact radial value through
\begin{equation}
\mathcal{E}_{N}
=
\frac{
\left|
\alpha^{(N)}
-
\alpha_{\mathrm{rad}}
\right|
}{
\alpha_{\mathrm{rad}}
}.
\end{equation}
Independent numerical evaluation of the exact radial integral in Eq. \eqref{6.8} and the exact phase integral gives agreement to the displayed numerical precision. Representative values of the exact bending angle and the relative truncation deviations are shown below.

\begin{table*}
\begin{tabular}{c|c|c|c|c|c}
$\epsilon$ &
$\alpha_{\mathrm{rad}}$ &
$100\mathcal{E}_{1}$ &
$100\mathcal{E}_{2}$ &
$100\mathcal{E}_{3}$ &
$100\mathcal{E}_{4}$ \\
\hline
0.01 &
0.041222539749274 &
2.965707 &
0.107811 &
0.00430793 &
0.000181841 \\
0.05 &
0.23613599538847 &
15.303044 &
2.830388 &
0.571802 &
0.121617 \\
0.10 &
0.59039578760583 &
32.248839 &
12.294475 &
5.067685 &
2.186773 \\
\end{tabular}
\caption{
Numerical values of the radiative coefficient and the scaled energies
$\mathcal{E}_{i}$ for selected values of $\epsilon$.}
\label{tab:6.44}
\end{table*}
The values in Table \ref{tab:6.44} show the expected asymptotic behavior. When \(\epsilon=0.01\), four terms reproduce the exact scattering angle to better than \(2\times10^{-6}\) in relative units. At \(\epsilon=0.05\), the fourth-order result remains accurate at the \(10^{-3}\) level. By \(\epsilon=0.10\), the same truncation still improves systematically with order, although the weak expansion converges much more slowly toward the exact value. The comparison therefore separates formal coefficient agreement from the practical usefulness of a low-order truncation.

\begin{figure*}[t]
\centering
\includegraphics[width=0.48\textwidth]{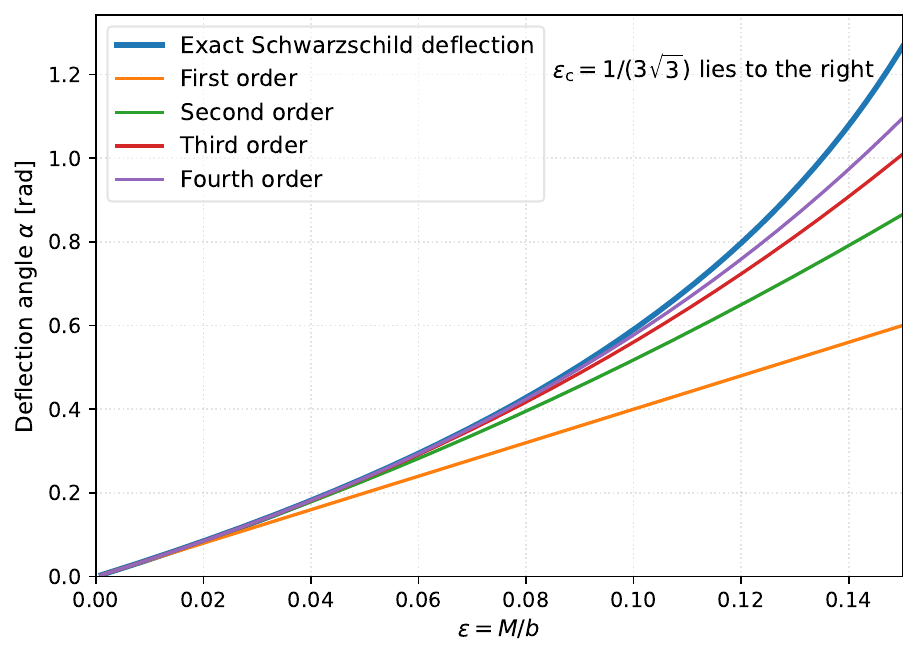}
\includegraphics[width=0.48\textwidth]{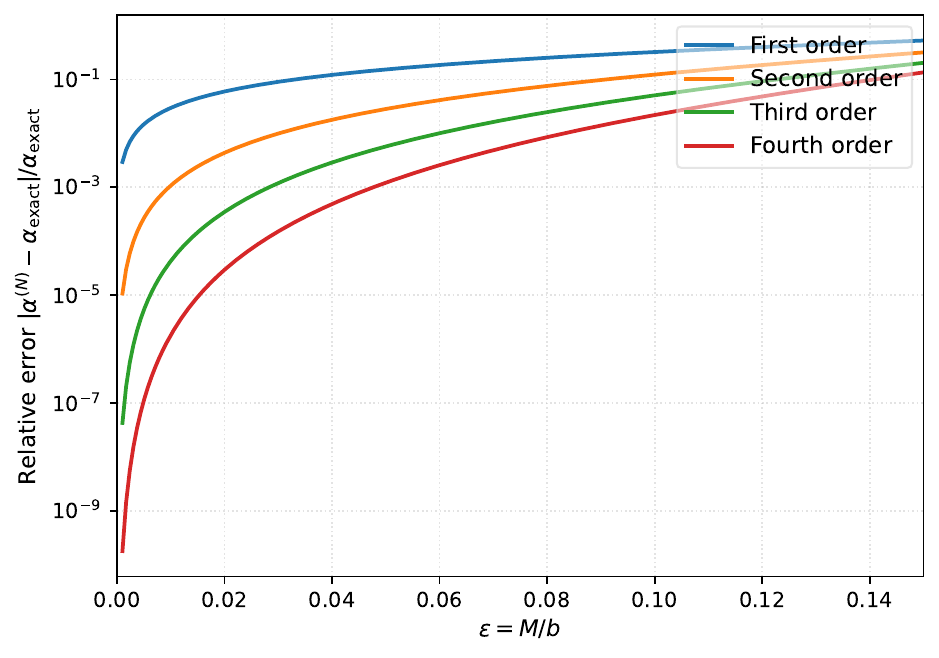}
\caption{Exact Schwarzschild bending and phase-series accuracy. The left panel compares the exact result with first through fourth-order truncations, while the right panel shows their relative errors.}
\label{fig4}
\end{figure*}
Successive truncations demonstrate (see Fig. \ref{fig4}) how the all-order phase expansion approaches the exact Schwarzschild result throughout the weak regime. The left panel shows that the first through fourth-order approximations become nearly indistinguishable from the exact curve when \(M/b\) is small, while their separation grows as the photon approaches the critical scattering scale. The right panel quantifies the same behavior and shows that each additional perturbative order reduces the relative error substantially when \(M/b\ll1\). The deterioration near the upper end of the plotted interval does not signal a failure of the exact phase formulation. It reflects the increasing number of terms required when \(M/b\) approaches the convergence boundary associated with the photon sphere. The two panels therefore distinguish the exact phase representation from the practical accuracy of a finite weak-field truncation.

A final exact check follows from the critical scattering boundary. The double-root analysis already carried out in Sec. \ref{sec2} applies directly to Eq. \eqref{6.2} and gives \(z_{\mathrm{c}}=\sqrt{3}\) and \(\epsilon_{\mathrm{c}}=1/(3\sqrt{3})\). The phase formulation reaches the same boundary independently through Eqs. \eqref{4.16} and \eqref{4.44}. The phase-factor denominator vanishes at \(y=1/\sqrt{3}\), for which the cubic gives \(q_{\mathrm{c}}=1/(3\sqrt{3})\). Since \(q=\epsilon\) at closest approach, the singular phase condition coincides exactly with the radial critical value.

The agreement is structurally significant. The radial formulation identifies the critical orbit through coalescence of turning points, while Eq. \eqref{4.35} shows that the phase formulation identifies it through \(d\Theta/d\phi\rightarrow0\). As the critical trajectory is approached, an increasingly large physical azimuth is required for a finite change in the intrinsic phase. The phase description therefore captures not only the weak-field coefficients but also the same obstruction that marks the transition toward the photon-sphere scattering regime.

Taken together, these checks establish equivalence at several levels. The exact radial integral and exact phase integral give the same scattering angle. The weak expansion reproduces the standard invariant Schwarzschild coefficients. The conventional orbit perturbation independently recovers the first two terms. Numerical evaluation confirms the behavior of successive truncations away from the strict \(\epsilon\rightarrow0\) limit. The algebraic phase map also identifies the same critical impact parameter as the radial turning-point analysis. We therefore retain the standard Schwarzschild observable while replacing the displaced-root calculation by a fixed-interval phase formulation.

\section{General Static Spherical Extension}
\label{sec7}
We now determine which parts of the phase-plane formulation belong specifically to Schwarzschild spacetime and which parts follow from static spherical symmetry alone. The distinction is essential. The fixed-phase description should have value beyond Schwarzschild only if we can derive it without relying on the particular cubic nonlinearity \(3\epsilon z^{2}\). We therefore return to the most general asymptotically flat static and spherically symmetric geometry and derive the phase equations directly from its null geodesics.

We begin with the line element
\begin{equation}
\begin{aligned}
ds^{2}
={}&
-\mathcal{A}(r)\,dt^{2}
+
\mathcal{B}(r)\,dr^{2}
\\
&+
\mathcal{C}(r)
\left(
d\theta^{2}
+
\sin^{2}\theta\,d\phi^{2}
\right).
\end{aligned}
\end{equation}
Asymptotic flatness requires
\begin{align}
\mathcal{A}(r)\rightarrow1,
\qquad
\mathcal{B}(r)\rightarrow1,
\qquad
\frac{\mathcal{C}(r)}{r^{2}}\rightarrow1
\end{align}
as \(r\rightarrow\infty\). We assume that the exterior scattering region admits a monotonically increasing areal radius. We therefore define \(R=\sqrt{\mathcal{C}(r)}\). The metric can then be written in the areal-radius form
\begin{equation}
\begin{aligned}
ds^{2}
={}&
-F(R)\,dt^{2}
+
G(R)\,dR^{2}
\\
&+
R^{2}
\left(
d\theta^{2}
+
\sin^{2}\theta\,d\phi^{2}
\right).
\end{aligned}
\end{equation}
The use of \(R\) is useful because the area of every symmetry sphere equals \(4\pi R^{2}\). We thereby remove an arbitrary radial parametrization before developing the weak-scattering expansion.

Spherical symmetry again allows us to choose the equatorial plane \(\theta = \pi/2\). Stationarity and axial symmetry give the conserved quantities
\begin{align}
E
=
F(R)\dot t,
\qquad
L
=
R^{2}\dot\phi.
\label{7.6}
\end{align}
We retain the invariant asymptotic impact parameter defined in Eq. \eqref{2.5}.

The null condition gives
\begin{equation}
-F(R)\dot t^{\,2}
+
G(R)\dot R^{\,2}
+
R^{2}\dot\phi^{\,2}
=
0.
\end{equation}
Using Eq. \eqref{7.6}, we obtain
\begin{equation}
G(R)\dot R^{\,2}
=
\frac{E^{2}}{F(R)}
-
\frac{L^{2}}{R^{2}}.
\end{equation}
Eliminating the affine parameter through \(\dot\phi=L/R^{2}\) gives
\begin{equation}
\left(
\frac{dR}{d\phi}
\right)^{2}
=
\frac{R^{4}}{G(R)}
\left[
\frac{1}{F(R)b^{2}}
-
\frac{1}{R^{2}}
\right].
\label{7.10}
\end{equation}

We introduce the dimensionless inverse areal radius \(z=b/R\). Since
\begin{equation}
\frac{dR}{d\phi}
=
-\frac{b}{z^{2}}
\frac{dz}{d\phi},
\end{equation}
Eq. \eqref{7.10} becomes
\begin{align}
z'^{\,2}
=
\frac{1}
{
F(R)G(R)
}
-
\frac{z^{2}}
{
G(R)
},
\end{align}
where \(R = b/z\). We denote the right-hand side by
\begin{equation}
\mathscr{H}(z)
=
\frac{1}
{
F(b/z)G(b/z)
}
-
\frac{z^{2}}
{
G(b/z)
}.
\label{7.14}
\end{equation}
Additional physical parameters carried by the metric remain implicit in \(\mathscr{H}\). For a one-scale black hole geometry characterized by the mass \(M\), the dependence on \(b\) can be expressed through the weak parameter \(\epsilon\) defined in Sec. \ref{sec2}. We then write \(\mathscr{H}=\mathscr{H}(z,\epsilon)\). Asymptotic flatness gives \(\mathscr{H}(0,\epsilon)=1\).

The complete radial scattering problem can now be represented by the first integral
\begin{equation}
z'^{\,2}
=
\mathscr{H}(z,\epsilon).
\label{7.18}
\end{equation}
Differentiating Eq. \eqref{7.18} with respect to \(\phi\) gives
\begin{equation}
2z'z''
=
\frac{\partial\mathscr{H}}{\partial z}
z'.
\end{equation}
Away from a turning point we obtain
\begin{equation}
z''
=
\frac{1}{2}
\frac{\partial\mathscr{H}}{\partial z}.
\end{equation}
Regularity extends this equation continuously through an ordinary turning point.

It is useful to isolate the flat-space harmonic term. We define
\begin{equation}
\mathcal{N}(z,\epsilon)
=
z
+
\frac{1}{2}
\frac{\partial\mathscr{H}}{\partial z}.
\label{7.21}
\end{equation}
The null orbit therefore satisfies
\begin{equation}
z''+z
=
\mathcal{N}(z,\epsilon).
\end{equation}
For Minkowski spacetime,
\begin{equation}
\mathscr{H}_{0}
=
1-z^{2},
\end{equation}
and Eq. \eqref{7.21} gives \(\mathcal{N}_{0}=0\). Hence \(\mathcal{N}\) measures the entire departure of the inverse-radius orbit from the flat harmonic equation.

We now use the same phase-plane variables defined in Eq. \eqref{3.2}. Repeating the exact algebra of Sec. \ref{sec3} gives
\begin{equation}
\frac{dA}{d\phi}
=
\cos\Theta\,
\mathcal{N}
\left(
A\sin\Theta,\epsilon
\right)
\label{7.26}
\end{equation}
and
\begin{equation}
\frac{d\Theta}{d\phi}
=
1
-
\frac{\sin\Theta}{A}
\mathcal{N}
\left(
A\sin\Theta,\epsilon
\right).
\label{7.27}
\end{equation}
No weak-field approximation has entered Eqs. \eqref{7.26} and \eqref{7.27}. They apply to every static spherical geometry whose null orbit can be represented by Eq. \eqref{7.18}.

The radial first integral provides an additional simplification that does not require us to solve Eq. \eqref{7.26}. From Eq. \eqref{3.2}, we have
\begin{equation}
A^{2}
=
z^{2}
+
z'^{\,2}.
\label{7.28}
\end{equation}
Using Eq. \eqref{7.18}, we find the exact implicit amplitude relation
\begin{equation}
A^{2}
=
z^{2}
+
\mathscr{H}(z,\epsilon).
\label{7.29}
\end{equation}
Since \(z=A\sin\Theta\), Eq. \eqref{7.29} becomes
\begin{equation}
A^{2}
=
A^{2}\sin^{2}\Theta
+
\mathscr{H}
\left(
A\sin\Theta,\epsilon
\right).
\end{equation}
Equivalently,
\begin{equation}
A^{2}\cos^{2}\Theta
=
\mathscr{H}
\left(
A\sin\Theta,\epsilon
\right).
\label{7.31}
\end{equation}
The amplitude is therefore determined locally and implicitly once the intrinsic phase is specified. For polynomial forms of \(\mathscr{H}\), Eq. \eqref{7.31} becomes an algebraic relation. For more general metric functions, the same equation remains an implicit local relation even when no elementary closed form exists.

We can eliminate the amplitude entirely if desired. Using
\begin{equation}
\sin^{2}\Theta
=
\frac{z^{2}}
{
z^{2}+\mathscr{H}(z,\epsilon)
},
\end{equation}
we obtain
\begin{equation}
\mathscr{H}(z,\epsilon)
=
z^{2}\cot^{2}\Theta.
\label{7.33}
\end{equation}
Equation \eqref{7.33} determines \(z\) locally as a function of \(\Theta\). At the incoming and outgoing asymptotes, \(z\rightarrow0\). At closest approach, \(\Theta = \pi/2\), so Eq. \eqref{7.33} reduces to \(\mathscr{H}(z_{\mathrm{m}},\epsilon) = 0\). The radial turning-point condition therefore appears automatically as the midpoint condition of the phase description.

The phase rate also admits a representation entirely in terms of \(\mathscr{H}\). From Eqs. \eqref{7.21} and \eqref{7.27}, together with \(z=A\sin\Theta\), we obtain
\begin{equation}
\frac{d\Theta}{d\phi}
=
1
-
\frac{z}
{
A^{2}
}
\left[
z
+
\frac{1}{2}
\frac{\partial\mathscr{H}}{\partial z}
\right].
\end{equation}
Using Eq. \eqref{7.29}, this expression simplifies to
\begin{equation}
\frac{d\Theta}{d\phi}
=
\frac{
\mathscr{H}(z,\epsilon)
-
\displaystyle
\frac{z}{2}
\frac{\partial\mathscr{H}}{\partial z}
}
{
z^{2}
+
\mathscr{H}(z,\epsilon)
}.
\label{7.37}
\end{equation}

We define
\begin{equation}
\mathscr{D}(z,\epsilon)
=
\mathscr{H}(z,\epsilon)
-
\frac{z}{2}
\frac{\partial\mathscr{H}}{\partial z}.
\label{7.38}
\end{equation}
The exact phase rate then reads
\begin{equation}
\frac{d\Theta}{d\phi}
=
\frac{
\mathscr{D}(z,\epsilon)
}
{
z^{2}
+
\mathscr{H}(z,\epsilon)
}.
\label{7.39}
\end{equation}
The denominator equals \(A^{2}\) and remains positive along a regular scattering trajectory. Hence monotonicity of the phase follows whenever \(\mathscr{D}(z,\epsilon)>0\) between the two asymptotes.

Under this condition, we invert Eq. \eqref{7.39} and obtain
\begin{equation}
\frac{d\phi}{d\Theta}
=
\frac{
z^{2}
+
\mathscr{H}(z,\epsilon)
}
{
\mathscr{D}(z,\epsilon)
}.
\end{equation}
The asymptotically flat incoming branch again starts at \(\Theta_{\mathrm{in}}=0\) while the outgoing branch terminates at \(\Theta_{\mathrm{out}}=\pi\). The exact bending angle therefore becomes
\begin{equation}
\alpha
=
\int_{0}^{\pi}
\left[
\frac{
z^{2}
+
\mathscr{H}(z,\epsilon)
}
{
\mathscr{D}(z,\epsilon)
}
-1
\right]
d\Theta,
\label{7.44}
\end{equation}
where \(z=z(\Theta)\) follows implicitly from Eq. \eqref{7.33}.

Equation \eqref{7.44} provides the general static spherical phase formula. The fixed integration interval does not depend on the detailed form of the metric. The metric enters through \(\mathscr{H}\), its derivative with respect to \(z\), and the local relation in Eq. \eqref{7.33}. We therefore retain the principal conceptual feature of the Schwarzschild formulation even when its exceptionally simple cubic relation is absent.

The same formula also clarifies the relation between phase monotonicity and unstable circular null motion. A critical returning trajectory occurs when the radial turning point becomes degenerate. In the present notation, the critical conditions are
\begin{equation}
\mathscr{H}
\left(
z_{\mathrm{c}},\epsilon_{\mathrm{c}}
\right)
=
0
\label{7.45}
\end{equation}
and
\begin{equation}
\left.
\frac{\partial\mathscr{H}}{\partial z}
\right|_{
z=z_{\mathrm{c}},
\epsilon=\epsilon_{\mathrm{c}}
}
=
0.
\label{7.46}
\end{equation}
Equations \eqref{7.38}, \eqref{7.45}, and \eqref{7.46} immediately give
\begin{equation}
\mathscr{D}
\left(
z_{\mathrm{c}},\epsilon_{\mathrm{c}}
\right)
=
0.
\end{equation}
Thus the intrinsic phase rate vanishes precisely when the outer scattering orbit reaches a degenerate radial turning point. The phase description and the usual photon-orbit condition encode the same critical geometry.

We next examine the weak-field structure in a form that makes comparison among different metrics straightforward. For a one-scale asymptotically flat geometry, we write
\begin{equation}
\mathscr{H}(z,\epsilon)
=
1-z^{2}
+
\sum_{n=1}^{\infty}
\epsilon^{n}h_{n}(z).
\end{equation}
The functions \(h_{n}(z)\) contain the metric-dependent corrections at each weak order. Equation \eqref{7.21} then gives
\begin{equation}
\mathcal{N}(z,\epsilon)
=
\frac{1}{2}
\sum_{n=1}^{\infty}
\epsilon^{n}
\frac{dh_{n}}{dz}.
\label{7.49}
\end{equation}
The cancellation of the flat harmonic term is exact.

The amplitude relation becomes
\begin{equation}
A^{2}
=
1
+
\sum_{n=1}^{\infty}
\epsilon^{n}
h_{n}
\left(
A\sin\Theta
\right).
\label{7.50}
\end{equation}
The phase rate follows from Eq. \eqref{7.37} as
\begin{equation}
\frac{d\Theta}{d\phi}
=
\frac{
1
+
\displaystyle
\sum_{n=1}^{\infty}
\epsilon^{n}
\left[
h_{n}(z)
-
\frac{z}{2}
h_{n}'(z)
\right]
}
{
1
+
\displaystyle
\sum_{n=1}^{\infty}
\epsilon^{n}
h_{n}(z)
}.
\label{7.51}
\end{equation}
Equations \eqref{7.50} and \eqref{7.51} give a direct perturbative scheme for any metric whose asymptotic functions admit a regular expansion in \(M/R\).

We can see the first-order content explicitly by writing
\begin{equation}
F(R)
=
1
+
a_{1}\frac{M}{R}
+
a_{2}
\left(
\frac{M}{R}
\right)^{2}
+
\mathcal{O}
\left[
\left(
\frac{M}{R}
\right)^{3}
\right]
\end{equation}
and
\begin{equation}
G(R)
=
1
+
\beta_{1}\frac{M}{R}
+
\beta_{2}
\left(
\frac{M}{R}
\right)^{2}
+
\mathcal{O}
\left[
\left(
\frac{M}{R}
\right)^{3}
\right].
\end{equation}
Substitution into Eq. \eqref{7.14} gives
\begin{equation}
\mathscr{H}
=
1-z^{2}
+
\epsilon
\left[
-
\left(
a_{1}+\beta_{1}
\right)z
+
\beta_{1}z^{3}
\right]
+
\mathcal{O}(\epsilon^{2}).
\end{equation}
Hence
\begin{equation}
h_{1}(z)
=
-
\left(
a_{1}+\beta_{1}
\right)z
+
\beta_{1}z^{3}.
\label{7.55}
\end{equation}
At leading order we may set
\begin{equation}
z
=
\sin\Theta
+
\mathcal{O}(\epsilon).
\end{equation}
Using Eq. \eqref{7.49}, the phase formula gives
\begin{equation}
\alpha
=
\frac{\epsilon}{2}
\int_{0}^{\pi}
\sin\Theta\,
h_{1}'
\left(
\sin\Theta
\right)
d\Theta
+
\mathcal{O}(\epsilon^{2}).
\label{7.57}
\end{equation}
From Eq. \eqref{7.55}, we obtain
\begin{equation}
h_{1}'(z)
=
-
\left(
a_{1}+\beta_{1}
\right)
+
3\beta_{1}z^{2}.
\end{equation}
Using the elementary moments \(\int_{0}^{\pi}\sin\Theta\,d\Theta=2\) and \(\int_{0}^{\pi}\sin^{3}\Theta\,d\Theta=4/3\), Eq. \eqref{7.57} becomes
\begin{equation}
\alpha
=
\left(
\beta_{1}-a_{1}
\right)
\frac{M}{b}
+
\mathcal{O}
\left[
\left(
\frac{M}{b}
\right)^{2}
\right].
\label{7.61}
\end{equation}
For Schwarzschild spacetime, \(a_{1}=-2\) and \(\beta_{1}=2\), so Eq. \eqref{7.61} immediately reproduces the leading result already stated in Sec. \ref{sec2}. The coefficient therefore follows directly from the first asymptotic terms of the temporal and radial metric functions.

We can sharpen the algebraic classification by asking when the amplitude equation becomes a finite polynomial relation. Suppose the orbit equation contains a single nonlinear monomial
\begin{equation}
z''+z
=
\lambda
\epsilon^{p}
z^{m},
\label{7.64}
\end{equation}
where \(m\geq1\) and \(p\geq1\). The corresponding first integral satisfying the asymptotic normalization is
\begin{equation}
z'^{\,2}
=
1-z^{2}
+
\frac{
2\lambda
}{
m+1
}
\epsilon^{p}
z^{m+1}.
\end{equation}
Using Eq. \eqref{7.28}, we obtain
\begin{equation}
A^{2}
=
1
+
\frac{
2\lambda
}{
m+1
}
\epsilon^{p}
A^{m+1}
\sin^{m+1}\Theta.
\label{7.66}
\end{equation}

We define
\begin{equation}
q
=
\epsilon^{p}
\sin^{m+1}\Theta.
\end{equation}
Equation \eqref{7.66} becomes
\begin{equation}
2\lambda qA^{m+1}
-
(m+1)A^{2}
+
(m+1)
=
0.
\label{7.68}
\end{equation}
Hence every single-monomial orbit equation of the form in Eq. \eqref{7.64} possesses an exact algebraic amplitude relation.

Introducing \(y=A^{-1}\), we rewrite Eq. \eqref{7.68} as
\begin{equation}
2\lambda q
=
(m+1)
y^{m-1}
\left(
1-y^{2}
\right).
\label{7.70}
\end{equation}
The exact phase equation follows directly from Eq. \eqref{7.27},
\begin{equation}
\frac{d\Theta}{d\phi}
=
1
-
\lambda qA^{m-1}.
\end{equation}
Using Eq. \eqref{7.70}, we obtain
\begin{equation}
\frac{d\Theta}{d\phi}
=
\frac{
(m+1)y^{2}
-
(m-1)
}{2}.
\label{7.72}
\end{equation}
Differentiation of Eq. \eqref{7.70} gives
\begin{equation}
\frac{dq}{dy}
=
\frac{
m+1
}{
2\lambda
}
y^{m-2}
\left[
(m-1)
-
(m+1)y^{2}
\right].
\label{7.73}
\end{equation}
Combining Eqs. \eqref{7.72} and \eqref{7.73}, we find
\begin{equation}
\frac{d\phi}{d\Theta}
=
-
\frac{
m+1
}{
\lambda
}
y^{m-2}
\frac{dy}{dq}.
\label{7.74}
\end{equation}
The Schwarzschild result follows by taking
\begin{align}
m=2,
\qquad
p=1,
\qquad
\lambda=3.
\end{align}
Equations \eqref{7.70} and \eqref{7.74} then reduce respectively to the Schwarzschild relations \eqref{4.16} and \eqref{4.39}. The cubic map is therefore one member of a wider algebraic family. Its exceptional simplicity follows from the quadratic orbit nonlinearity, which removes the additional power of \(y\) from Eq. \eqref{7.74}.

The same reasoning extends to finite sums of monomials. Suppose
\begin{equation}
z''+z
=
\sum_{k=1}^{K}
\lambda_{k}
\epsilon^{p_{k}}
z^{m_{k}}.
\end{equation}
Integration of the orbit equation gives
\begin{equation}
z'^{\,2}
=
1-z^{2}
+
\sum_{k=1}^{K}
\frac{
2\lambda_{k}
}{
m_{k}+1
}
\epsilon^{p_{k}}
z^{m_{k}+1}.
\end{equation}
The phase-plane amplitude then obeys
\begin{equation}
A^{2}
=
1
+
\sum_{k=1}^{K}
\frac{
2\lambda_{k}
}{
m_{k}+1
}
\epsilon^{p_{k}}
A^{m_{k}+1}
\sin^{m_{k}+1}\Theta.
\label{7.80}
\end{equation}
Equation \eqref{7.80} is algebraic in \(A\) whenever the sum contains finitely many integer powers. We therefore do not need a differential amplitude equation even when several nonlinear terms occur. What generally disappears is the reduction to one cubic inverse map with one composite variable. The Schwarzschild case remains unusually compact because its exact orbit equation contains only one nonlinear monomial.

A charged nonrotating black hole provides a useful illustration. For the Reissner--Nordstr\"om geometry, we have
\begin{align}
F(R)
&=
1
-
\frac{2M}{R}
+
\frac{Q^{2}}{R^{2}},
\nn
G(R)
&=
\frac{1}{F(R)}.
\end{align}
Using the previously defined \(\epsilon\), we introduce the additional charge parameter \(\eta=Q^{2}/b^{2}\).
Since \(F(R)G(R)=1\), Eq. \eqref{7.14} gives
\begin{equation}
\mathscr{H}(z)
=
1-z^{2}
+
2\epsilon z^{3}
-
\eta z^{4}.
\end{equation}
The corresponding orbit equation becomes
\begin{equation}
z''+z
=
3\epsilon z^{2}
-
2\eta z^{3}.
\end{equation}
The phase-plane amplitude satisfies the exact quartic relation
\begin{equation}
A^{2}
=
1
+
2\epsilon
A^{3}\sin^{3}\Theta
-
\eta
A^{4}\sin^{4}\Theta.
\label{7.85}
\end{equation}
The exact phase rate is
\begin{equation}
\frac{d\Theta}{d\phi}
=
1
-
3\epsilon
A\sin^{3}\Theta
+
2\eta
A^{2}\sin^{4}\Theta.
\label{7.86}
\end{equation}
Thus the Schwarzschild cubic relation becomes a quartic relation when the charge term is present, while the fixed phase interval remains unchanged.

To compare the weak expansion with the familiar charged result, we hold the dimensionless ratio \(\chi = Q/M\) fixed. We then have \(\eta = \chi^{2}\epsilon^{2}\). Equation \eqref{7.85} gives the leading amplitude
\begin{equation}
A
=
1
+
\epsilon\sin^{3}\Theta
+
\mathcal{O}(\epsilon^{2}).
\end{equation}
Substitution into Eq. \eqref{7.86}, followed by expansion of its reciprocal, gives
\begin{equation}
\begin{aligned}
\frac{d\phi}{d\Theta}
-
1
={}&
3\epsilon\sin^{3}\Theta
+
\epsilon^{2}
\left[
12\sin^{6}\Theta
-
2\chi^{2}\sin^{4}\Theta
\right]
\\
&+
\mathcal{O}(\epsilon^{3}).
\end{aligned}
\end{equation}
The required phase moments are
\begin{equation}
\int_{0}^{\pi}
\sin^{4}\Theta\,d\Theta
=
\frac{3\pi}{8}
\end{equation}
and
\begin{equation}
\int_{0}^{\pi}
\sin^{6}\Theta\,d\Theta
=
\frac{5\pi}{16}.
\end{equation}
We therefore obtain
\begin{equation}
\alpha
=
4\epsilon
+
\left(
\frac{15\pi}{4}
-
\frac{3\pi}{4}\chi^{2}
\right)
\epsilon^{2}
+
\mathcal{O}(\epsilon^{3}).
\end{equation}
Restoring \(M\), \(Q\), and \(b\), we find
\begin{equation}
\alpha
=
\frac{4M}{b}
+
\frac{
15\pi M^{2}
-
3\pi Q^{2}
}{
4b^{2}
}
+
\mathcal{O}
\left(
\frac{M^{3}}{b^{3}}
\right)
\end{equation}
for fixed \(Q/M\). The phase framework therefore reproduces the standard charged correction while preserving the same fixed interval \(0\leq\Theta\leq\pi\).

The Reissner--Nordstr\"om example also clarifies what we mean by generalization. We do not require every spacetime to reproduce the Schwarzschild cubic map. The more fundamental statement is that the first radial integral determines the phase amplitude locally through Eq. \eqref{7.29}, while the scattering angle follows from the fixed-phase integral in Eq. \eqref{7.44}. Polynomial radial first integrals yield finite algebraic relations for \(A\). A single monomial yields the particularly compact family in Eqs. \eqref{7.68}--\eqref{7.74}.

We can therefore separate three levels of structure. Static spherical symmetry and asymptotic flatness give the invariant first integral in Eq. \eqref{7.18}. The phase transformation converts that first integral into the fixed-interval scattering formula in Eq. \eqref{7.44}. Additional simplicity arises when \(\mathscr{H}\) has a finite polynomial dependence on \(z\), because the amplitude then follows from a local algebraic equation rather than from differential evolution.

The Schwarzschild geometry occupies the simplest nontrivial point in this hierarchy because its radial function \(\mathscr{H}_{\mathrm{Sch}}=1-z^{2}+2\epsilon z^{3}\) contains only one nonlinear correction. Consequently, its amplitude relation reduces to one cubic equation, its phase factor reduces to the derivative of one inverse cubic map, and its all-order weak coefficients factor into one algebraic sequence and one family of trigonometric moments. More general static spherical metrics retain the phase formulation while replacing this single cubic structure by the corresponding implicit relation generated by their radial geodesic function.

The extension developed here therefore shows that the phase-plane method is not tied to a fortunate rewriting of the Schwarzschild equation. Its essential content follows from the first-integral structure of static spherical null motion. Schwarzschild spacetime is special because that general structure becomes exceptionally simple, not because the phase formulation itself ceases to exist outside Schwarzschild geometry.

\section{Validity Domain and Mathematical Limitations}
\label{sec8}
We now identify the precise domain in which the phase formulation and its weak-deflection series apply. Several logically different statements must be separated. The phase representation itself is exact for every returning Schwarzschild null orbit outside the critical trajectory. The power series derived in Sec. \ref{sec5} requires analyticity of the physical algebraic branch. The usefulness of a finite truncation imposes a stronger practical restriction because convergence becomes increasingly slow as the photon sphere is approached.

For Schwarzschild scattering, a returning orbit exists when \(0<\epsilon<\epsilon_{\mathrm{c}}\), where \(\epsilon_{\mathrm{c}}=1/(3\sqrt{3})\). Equivalently,
\begin{align}
b
>
b_{\mathrm{c}},
\qquad
b_{\mathrm{c}}
=
3\sqrt{3}\,M.
\label{8.4}
\end{align}
The limiting value in Eq. \eqref{8.4} corresponds to the unstable circular null orbit at \(r=3M\).

The exact phase description uses the composite variable in Eq. \eqref{4.4}, the reciprocal amplitude \(y=1/A\), and the cubic map \eqref{4.16}. The physical branch satisfies \(y(0)=1\) and remains within
\begin{equation}
\frac{1}{\sqrt{3}}
<
y
\leq
1
\label{8.7}
\end{equation}
for every returning orbit. The phase factor and exact bending angle are given by Eqs. \eqref{4.44} and \eqref{4.46}, respectively. No assumption of small \(\epsilon\) is required in Eq. \eqref{4.46}; the essential restrictions are regularity of the physical branch and positive intrinsic phase advance. Equation \eqref{4.35}, together with the bound in Eq. \eqref{8.7}, guarantees \(d\Theta/d\phi>0\). The fixed interval \(0\leq\Theta\leq\pi\) therefore remains a valid parametrization throughout the complete returning trajectory.

The weak-deflection series \eqref{5.13} introduces an additional mathematical requirement. Its convergence is determined by the nearest singularity of the inverse algebraic branch \(y(q)\). From Eq. \eqref{4.16}, the inverse ceases to remain locally regular when
\begin{equation}
\frac{dq}{dy}
=
0.
\end{equation}
Differentiation gives
\begin{equation}
\frac{dq}{dy}
=
\frac{
1-3y^{2}
}{2}.
\end{equation}
The branch point connected to the physical scattering branch therefore occurs at \(y_{\mathrm{c}}=1/\sqrt{3}\). Substitution into Eq. \eqref{4.16} gives \(q_{\mathrm{c}}=1/(3\sqrt{3})\).

The equality \(q_{\mathrm{c}} =\epsilon_{\mathrm{c}} \) is not accidental. At closest approach, \(\Theta = \pi/2\), so that \(q_{\max} = \epsilon \). The physical phase trajectory reaches the algebraic branch point precisely when the impact parameter reaches the critical Schwarzschild value.

For every fixed \(0\leq\epsilon<\epsilon_{\mathrm{c}}\), we have \(0\leq q\leq\epsilon<q_{\mathrm{c}}\). The Taylor series \eqref{5.13} consequently converges uniformly over the full phase interval, justifying the termwise integration that leads to Eqs. \eqref{5.36} and \eqref{5.37}. The resulting series is therefore more than a formal expansion about \(\epsilon=0\): it converges to the exact phase integral throughout
\begin{equation}
\left|
\epsilon
\right|
<
\frac{1}{3\sqrt{3}}
\label{8.25}
\end{equation}
on the analytic branch connected to weak Schwarzschild scattering.

Physical applications in the present work use only positive \(\epsilon\). The complex extension in Eq. \eqref{8.25} serves to characterize the Taylor series rather than to assign a scattering interpretation to negative or complex mass parameters.

We can understand the loss of analyticity near the critical orbit directly from the cubic map. We write
\begin{align}
y
=
y_{\mathrm{c}}
+
x,
\qquad
y_{\mathrm{c}}
=
\frac{1}{\sqrt{3}}.
\end{align}
Expansion of Eq. \eqref{4.16} about \(x=0\) gives
\begin{equation}
q
=
q_{\mathrm{c}}
-
\frac{\sqrt{3}}{2}x^{2}
-
\frac{1}{2}x^{3}.
\end{equation}
The leading behavior therefore satisfies
\begin{equation}
x
=
\left[
\frac{
2
\left(
q_{\mathrm{c}}-q
\right)
}
{
\sqrt{3}
}
\right]^{1/2}
+
\mathcal{O}
\left(
q_{\mathrm{c}}-q
\right).
\label{8.28}
\end{equation}
The phase factor in Eq. \eqref{4.44} becomes
\begin{equation}
\mathcal{F}
=
\frac{1}
{
\sqrt{3}\,x
}
-
\frac{1}{2}
+
\mathcal{O}(x).
\label{8.29}
\end{equation}
Combining Eqs. \eqref{8.28} and \eqref{8.29}, we obtain
\begin{equation}
\mathcal{F}(q)
=
\frac{
1
}{
\sqrt{2}\,3^{1/4}
}
\frac{
1
}{
\sqrt{
q_{\mathrm{c}}-q
}
}
-
\frac{1}{2}
+
\mathcal{O}
\left[
\sqrt{
q_{\mathrm{c}}-q
}
\right].
\label{8.30}
\end{equation}
Thus the algebraic phase factor develops a square-root singularity as the critical orbit is approached.

The singularity acquires a stronger form after the phase integration because \(q\) reaches its maximum at \(\Theta=\pi/2\). We define \(\xi = \Theta - \pi/2\) and \(\Delta = q_{\mathrm{c}}-\epsilon\). Near closest approach,
\begin{equation}
\sin^{3}\Theta
=
1
-
\frac{3}{2}\xi^{2}
+
\mathcal{O}(\xi^{4}).
\end{equation}
Hence
\begin{equation}
q_{\mathrm{c}}
-
q
=
\Delta
+
\frac{3\epsilon}{2}\xi^{2}
+
\mathcal{O}(\xi^{4}).
\end{equation}
Taking the limit \(\epsilon\rightarrow q_{\mathrm{c}}\) in the coefficient of \(\xi^{2}\) gives
\begin{equation}
q_{\mathrm{c}}
-
q
=
\Delta
+
\frac{
\xi^{2}
}{
2\sqrt{3}
}
+
\mathcal{O}
\left(
\Delta\xi^{2}
\right)
+
\mathcal{O}(\xi^{4}).
\end{equation}

Substitution into Eq. \eqref{8.30} gives the leading critical phase integrand
\begin{equation}
\mathcal{F}
\left(
\epsilon\sin^{3}\Theta
\right)
=
\frac{
1
}{
\sqrt{
\xi^{2}
+
2\sqrt{3}\,\Delta
}
}
+
\mathcal{O}(1).
\end{equation}
The singular part of the bending integral therefore contains
\begin{equation}
\int_{-\xi_{0}}^{\xi_{0}}
\frac{
d\xi
}{
\sqrt{
\xi^{2}
+
2\sqrt{3}\,\Delta
}
},
\end{equation}
where \(\xi_{0}\) is fixed and sufficiently small. Elementary integration gives
\begin{equation}
\int_{-\xi_{0}}^{\xi_{0}}
\frac{
d\xi
}{
\sqrt{
\xi^{2}
+
2\sqrt{3}\,\Delta
}
}
=
2
\operatorname{arsinh}
\left[
\frac{
\xi_{0}
}{
\sqrt{
2\sqrt{3}\,\Delta
}
}
\right].
\end{equation}
As \(\Delta\rightarrow0^{+}\), we obtain
\begin{equation}
\alpha
=
-\ln\Delta
+
\mathcal{O}(1).
\label{8.39}
\end{equation}

We can rewrite the critical behavior in terms of the invariant impact parameter. Since \(\epsilon=M/b\) and \(q_{\mathrm{c}}=M/b_{\mathrm{c}}\), we define \(\delta_{b}=(b/b_{\mathrm{c}})-1\). For \(\delta_{b}\ll1\),
\begin{equation}
q_{\mathrm{c}}-\epsilon
=
q_{\mathrm{c}}
\delta_{b}
+
\mathcal{O}(\delta_{b}^{2}).
\end{equation}
Equation \eqref{8.39} then gives
\begin{equation}
\alpha
=
-
\ln
\left(
\frac{b}{b_{\mathrm{c}}}-1
\right)
+
\mathcal{O}(1).
\label{8.44}
\end{equation}
The phase formulation therefore recovers the unit coefficient multiplying the Schwarzschild logarithmic divergence. The weak-series branch point and the strong-deflection photon-sphere singularity are two manifestations of the same algebraic structure.

This relation also determines the large-order behavior of the weak coefficients. Equation \eqref{8.30} can be written as
\begin{equation}
\mathcal{F}(q)
=
\sqrt{
\frac{3}{2}
}
\left(
1-\frac{q}{q_{\mathrm{c}}}
\right)^{-1/2}
+
\mathcal{O}(1)
\end{equation}
near the dominant positive singularity. Standard coefficient asymptotics then gives
\begin{equation}
f_{n}
=
\sqrt{
\frac{3}{2\pi n}
}
q_{\mathrm{c}}^{-n}
\left[
1
+
\mathcal{O}
\left(
\frac{1}{n}
\right)
\right].
\label{8.46}
\end{equation}
The trigonometric moment entering Eq. \eqref{5.37} satisfies
\begin{equation}
\frac{
\sqrt{\pi}
\Gamma
\left(
\frac{3n+1}{2}
\right)
}
{
\Gamma
\left(
\frac{3n+2}{2}
\right)
}
=
\sqrt{
\frac{
2\pi
}{
3n
}
}
\left[
1
+
\mathcal{O}
\left(
\frac{1}{n}
\right)
\right].
\label{8.47}
\end{equation}
Combining Eqs. \eqref{8.46} and \eqref{8.47}, we obtain the remarkably simple result
\begin{equation}
\mathcal{A}_{n}
=
\frac{
q_{\mathrm{c}}^{-n}
}{
n
}
\left[
1
+
\mathcal{O}
\left(
\frac{1}{n}
\right)
\right].
\end{equation}
Using \(q_{\mathrm{c}}^{-1}=3\sqrt{3}\), we may write
\begin{equation}
\mathcal{A}_{n}
=
\frac{
\left(
3\sqrt{3}
\right)^{n}
}{
n
}
\left[
1
+
\mathcal{O}
\left(
\frac{1}{n}
\right)
\right].
\label{8.49}
\end{equation}
At the critical value \(\epsilon=q_{\mathrm{c}}\), the \(n\)-th term therefore behaves as
\begin{equation}
\mathcal{A}_{n}
\epsilon_{\mathrm{c}}^{n}
=
\frac{1}{n}
\left[
1
+
\mathcal{O}
\left(
\frac{1}{n}
\right)
\right].
\end{equation}
The harmonic large-order behavior explains the logarithmic divergence in Eq. \eqref{8.44} directly from the weak-field coefficients.

Although the infinite weak series converges for every \(\epsilon<\epsilon_{\mathrm{c}}\), a finite truncation becomes progressively less efficient when \(\epsilon\) approaches \(\epsilon_{\mathrm{c}}\). Equation \eqref{8.49} shows why. The magnitude of the \(n\)-th contribution behaves at large \(n\) as
\begin{equation}
\mathcal{A}_{n}\epsilon^{n}
=
\frac{1}{n}
\left(
\frac{
\epsilon
}{
\epsilon_{\mathrm{c}}
}
\right)^{n}
\left[
1
+
\mathcal{O}
\left(
\frac{1}{n}
\right)
\right].
\end{equation}
When \(\epsilon/\epsilon_{\mathrm{c}}\) is small, successive terms decrease rapidly. When this ratio approaches unity, many orders are required before a truncated series approximates the exact bending angle accurately. Weak deflection should therefore refer to the usefulness of a low-order expansion rather than merely to the existence of the convergent infinite series.

The present formulation also assumes that both asymptotic endpoints lie at spatial infinity. We define the deflection angle through the difference between the total asymptotic azimuthal sweep and \(\pi\). If the source or observer lies at finite radius, the endpoint phases need not coincide with \(0\) and \(\pi\), and the observable angle must be related to local propagation directions measured by the corresponding observers. The same phase variables may still be useful, but Eq. \eqref{4.46} cannot be applied without modifying the endpoint geometry.

A separate limitation concerns the choice of radial variable. In Schwarzschild spacetime we define \(z=b/r\) using the areal radius \(r\). The phase-plane pair \((z,z')\), the amplitude \(A\), and the local phase \(\Theta\) depend on this representation. They are not local spacetime scalars. The invariant quantities are the asymptotic impact parameter \(b=L/E\) and the final scattering angle \(\alpha\).

For a general static spherical geometry, we reduced the radial freedom by choosing the areal radius \(R=\sqrt{\mathcal{C}}\). This choice gives a geometrically defined radial variable whenever the relevant exterior region admits a monotonic areal radius. If one instead introduces a different radial parametrization, the local amplitude and phase equations change. The integrated scattering angle must remain unchanged when the transformation is performed consistently.

We should therefore interpret the phase lag carefully. The local quantity \(d\phi/d\Theta - 1\) provides a useful mathematical representation of gravitational deflection within the adopted phase convention. We do not interpret it as a separately measurable local observable. Its integral between the physical asymptotic endpoints gives the invariant scattering angle.

The extension developed in Sec. \ref{sec7} introduces further restrictions. For a general static spherical metric, the exact phase representation requires a returning trajectory with one ordinary radial minimum and a monotonic intrinsic phase. In terms of the radial function \(\mathscr{H}(z,\epsilon)\), we require \(\mathscr{H}\left(z_{\mathrm{m}},\epsilon\right)=0 \) at the turning point and \(\mathscr{H}(z,\epsilon)>0\) along the open scattering branches. The phase rate remains monotonic when
\begin{equation}
\mathscr{D}(z,\epsilon)
=
\mathscr{H}(z,\epsilon)
-
\frac{z}{2}
\frac{
\partial\mathscr{H}
}{
\partial z
}
>
0.
\end{equation}
Metrics with several accessible turning points, nonmonotonic areal radius, or additional singular structures may require a piecewise phase description.

The algebraic simplification found for Schwarzschild spacetime is still more restrictive. The fixed-phase formulation follows quite generally, but the cubic reduction \eqref{4.16} depends on the single quadratic nonlinearity of the Schwarzschild orbit equation. Other geometries may yield higher-degree algebraic relations or implicit equations that cannot be inverted in elementary functions. The general phase method therefore survives more broadly than the particularly compact Schwarzschild coefficient formula.

Our conclusions concerning convergence also belong specifically to the Schwarzschild algebraic branch analyzed above. A different static spherical geometry can possess additional photon spheres or complex singularities closer to the weak-field expansion point. Its radius of convergence must therefore be determined from its own phase relation rather than inferred from Eq. \eqref{8.25}.

Within these restrictions, Eq. \eqref{4.46} gives an exact fixed-phase representation for every returning Schwarzschild orbit with \(b>b_{\mathrm{c}}\). The power series in \(M/b\) converges throughout the open interval \(0\leq M/b<1/(3\sqrt{3})\). Its limiting singularity coincides with the unstable photon orbit. The same singularity determines the large-order coefficient growth and produces the logarithmic divergence of the exact bending angle as \(b\rightarrow b_{\mathrm{c}}^{+}\). The weak and strong scattering limits therefore meet naturally within one phase-plane structure.

\section{Conclusions}
\label{sec9}
We have developed a phase-plane formulation of weak gravitational deflection for null rays in static and spherically symmetric spacetimes. Our central change of viewpoint is simple. Instead of determining the bending angle from the displacement of the outgoing zero of a perturbed orbit, we describe the null trajectory by an amplitude and an intrinsic phase and identify the deflection angle with the excess physical azimuth accumulated while the intrinsic phase advances through one fixed half-cycle. The incoming and outgoing asymptotes then correspond to fixed phase endpoints. Gravity changes the rate at which physical azimuth accumulates between them.

For Schwarzschild spacetime, this reformulation leads to a stronger simplification. The amplitude does not require an independent differential evolution once we combine the phase variables with the exact radial first integral. The amplitude becomes determined by a cubic algebraic relation through a composite phase variable. After introducing the reciprocal amplitude, the physical branch becomes the inverse of a simple cubic map. The local phase factor follows directly from the derivative of that inverse map. In this way, the original nonlinear second-order geodesic problem reduces to a local algebraic inversion followed by an integration over a fixed phase interval.

The weak-deflection expansion then acquires a transparent all-order structure. We generated the reciprocal-amplitude series through Lagrange inversion and obtained an explicit finite expression for the phase coefficients at arbitrary perturbative order. Each Schwarzschild bending coefficient separates into two factors. One factor carries the nonlinear information encoded by the algebraic phase map. The second factor is a universal trigonometric moment associated with the fixed phase interval. This separation explains why even powers of \(M/b\) contain a factor of \(\pi\), while odd powers remain rational. The familiar Schwarzschild coefficients arise directly from the same coefficient rule without solving a new orbit equation at each order.

Several independent checks establish that the reformulation preserves the standard physical scattering angle. The phase series reproduces the conventional weak Schwarzschild coefficients obtained from the exact radial integral. A separate orbit perturbation recovers the same first and second order terms through the displaced-root method. Numerical comparison with the exact radial scattering integral shows systematic improvement as successive phase terms are retained within the weak regime. The phase description also identifies the same critical impact parameter as the turning-point analysis of the radial equation.

The general static spherical extension shows that the fixed-phase formulation does not depend on the particular Schwarzschild cubic. Once we use the areal radius and the invariant impact parameter, the null radial first integral determines a local phase amplitude relation for any asymptotically flat static spherical geometry with an ordinary returning orbit. The bending angle can then be expressed over the same fixed intrinsic phase interval. Polynomial radial first integrals produce algebraic amplitude relations, while more general metrics lead to implicit local relations. Schwarzschild spacetime remains exceptional because its orbit equation contains one nonlinear monomial and therefore reduces to a single cubic inverse map.

The analysis of the admissible domain also reveals a direct relation between weak and critical scattering. The physical Schwarzschild algebraic branch remains regular for every returning orbit outside the photon sphere. Its first branch singularity occurs at the critical impact parameter. The same singularity governs the radius of convergence of the weak series, fixes the large-order growth of its coefficients, and produces the logarithmic divergence of the exact bending angle as the critical orbit is approached. Weak and strong Schwarzschild scattering therefore emerge from different regimes of the same phase-plane structure rather than from unrelated mathematical descriptions.

The method has clear limitations. The local amplitude and intrinsic phase depend on the chosen radial representation even though the final asymptotic scattering angle does not. The present formulation assumes an asymptotically flat trajectory with one incoming branch, one ordinary turning point, and one outgoing branch. Finite-distance source and observer configurations require modified endpoint conditions. More complicated static spherical geometries can also contain several turning points or several photon spheres, in which case a single global phase branch may no longer suffice. The compact all-order Schwarzschild coefficient formula should therefore be distinguished from the more general fixed-phase principle on which it rests.

Future work should first classify static spherical orbit equations according to whether their phase amplitudes admit finite algebraic relations and whether those relations possess coefficient formulas comparable to the Schwarzschild case. We should then examine finite-distance deflection within the same phase language and determine how local observational angles modify the fixed-endpoint picture. Another important direction is the analytic study of phase-map singularities for geometries with several unstable null orbits, since their location should determine the convergence properties of the corresponding weak series. We may also investigate whether the phase formulation can connect systematically with strong-deflection expansions through analytic continuation or rational approximations. These extensions would determine whether the present Schwarzschild result represents a special calculational simplification or the first member of a broader phase-based theory of gravitational scattering.

\begin{acknowledgments}
R. P. and A. \"O. would like to acknowledge networking support of the COST Action CA21106 - COSMIC WISPers in the Dark Universe: Theory, astrophysics and experiments (CosmicWISPers), the COST Action CA22113 - Fundamental challenges in theoretical physics (THEORY-CHALLENGES), the COST Action CA21136 - Addressing observational tensions in cosmology with systematics and fundamental physics (CosmoVerse), the COST Action CA23130 - Bridging high and low energies in search of quantum gravity (BridgeQG), and the COST Action CA23115 - Relativistic Quantum Information (RQI) funded by COST (European Cooperation in Science and Technology). R. P. and A. \"O. would also like to acknowledge the funding support of SCOAP3. A. \"O. also thanks to EMU, TUBITAK, ULAKBIM (Turkiye)
\end{acknowledgments}

\bibliography{ref}

\end{document}